\documentclass[11pt,a4paper]{article}
\pdfoutput=1
\usepackage{graphicx}
\usepackage{jheppub}

\usepackage{epstopdf}
\usepackage{amsfonts}
\usepackage{latexsym}
\usepackage{verbatim}
\usepackage{subfigure}
\usepackage{slashed}
\usepackage{dsfont}
\usepackage{natbib}
\newcommand{\m}{\mu}
\newcommand{\g}{\gamma}
\newcommand{\dau}{\partial}

\newcommand{\kv}{\vec{k}}
\newcommand{\psib}{\bar{\psi}}
\newcommand{\Tr}{\text{Tr}}
\newcommand{\tr}{\text{tr}}
\newcommand{\unit}{\mathds{1}}

\title{Towards a field-theory interpretation of bottom-up holography}
\author{V.P.J. Jacobs,}
\author{S. Grubinskas, }
\author{H.T.C. Stoof}
\affiliation{Institute for Theoretical Physics and Center for Extreme Matter and Emergent Phenomena,
Utrecht University, Leuvenlaan 4, 3584 CE Utrecht, The Netherlands
}
\emailAdd{V.P.J.Jacobs@uu.nl}
\emailAdd{S.Grubinskas@uu.nl}
\emailAdd{H.T.C.stoof@uu.nl}
\abstract{We investigate recent results for the electrical conductivity and the fermionic self-energy, obtained in a holographic bottom-up model for a relativistic charge-neutral conformal field theory. We present two possible field-theoretic derivations of these results, using either a semiholographic or a holographic point of view. In the semiholographic interpretation, we also show how, in general, the conductivity should be calculated in agreement with Ward identities. The resulting field-theory interpretation may lead to a better understanding of the holographic dictionary in applied AdS/CMT.}

\begin{document}

\DeclareGraphicsExtensions{.jpg,.JPG,.jpeg,.pdf,.png,.eps}
\graphicspath{{./graphics/}}

\maketitle

\section{Introduction}

A phase transition takes place when the free energy of a system has a nonanalyticity as a function of another thermodynamic variable of the system, such as the temperature or the pressure.
Phase transitions can be classified by the way the free energy changes at the transition, distinguishing discontinuous (first-order) and continuous (second or higher-order) phase transitions. During a first-order phase transition, the derivative of the free energy is discontinuous, and the variable describing the amount of order in the system, the order parameter, jumps from a zero to a nonzero value. In contrast, during a continuous phase transition the order parameter becomes nonzero in a continuous manner. In the latter case, precisely at this so-called critical point the system has a diverging correlation length. This implies that the correlations in the system look the same at all scales.

Consider, for instance, a system of fermions and bosons in the vicinity of such a continuous phase transition. This is a generic situation in experimental and theoretical condensed-matter physics, see for example Ref.~\cite{Wolfle07} and references therein. The bosons represent fluctuations in a collective field, the expectation value of which is the order parameter introduced above. As mentioned previously, there exists a control parameter $p$ which can be tuned to a critical value $p_c$, at which the phase transition occurs. There, the correlation length of the bosonic fluctuations becomes infinite and the bosonic degree of freedom obtains a nonzero expectation value at one side of the transition. Then the system is said to be in the ordered phase. Examples of such an ordered phase are a superfluid, an (anti)ferromagnet or a charge-density wave. When the system is still in the disordered phase, but close to the critical point, the expectation value of the order parameter is zero but there are critical bosonic fluctuations which have an increasing importance as $p\rightarrow p_c$.

\begin{figure}[t]
\vskip 10pt
\centering
\includegraphics[width=0.95\textwidth]{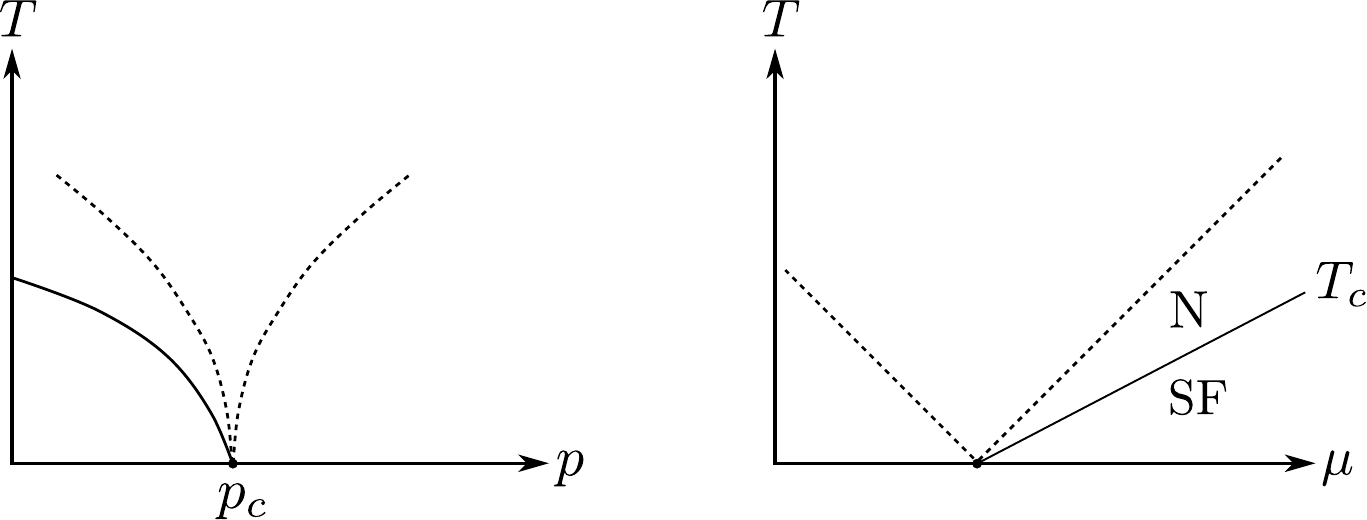}
\caption{Left panel: Phase diagram containing a quantum critical point at the $T=0$ end point of a line of phase transitions to an ordered phase. The dotted lines represent a cross-over to the quantum critical region which is directly above the quantum critical point.
Right panel: Phase diagram of ultracold fermionic atoms at unitarity. Above the critical temperature $T_c$, the system is in the disordered or normal state, denoted by ``N''. For positive values of the chemical potential $\mu$ there is a line of second-order phase transitions to the ordered superfluid phase (``SF''). These phase transitions can for $T>0$ be described using classical physics, but the line terminates in a quantum critical point at zero temperature and zero chemical potential. The dotted lines indicate the crossover from the classical regime to the quantum critical region, where the behavior is dominated by the physics of the quantum critical point.}
\label{fig:phasecrit}
\end{figure}

 A phase transition occurring at zero temperature is known as a quantum phase transition \cite{sondhi97,SachdevQPT,vojta03}, since the nature of the fluctuations of the order parameter is purely quantum. As suggested above, in the absence of a temperature scale and in the case of an infinite correlation length, the behavior of the order parameter fluctuations can become fully scale invariant
 precisely at the transition. This can be the case both for weakly and strongly coupled systems.
 A typical phase diagram containing a quantum critical point is shown in the left panel of Fig.~\ref{fig:phasecrit}. In this figure, the quantum critical point is the end point at $T=0$ of a line of nonzero-temperature phase transitions to an ordered phase. Interestingly, due to the absence of other scales, the quantum critical point dominates the behavior of the system at $p_c$ even for temperatures $T>0$, which is therefore called the quantum critical region.

We can ask what the consequences of this criticality are for the fermions that are also present in the system. In many cases, this can be studied using a field-theoretic description where the order-parameter fluctuations give rise to an effective interaction for the fermions. In the simplest approach, the bosonic system becomes scale invariant and can be modeled by a conformal field theory.\footnote{Conformal invariance is more restrictive than scale invariance as it also includes symmetry under special conformal transformations. What is most relevant for our purposes, is that the theory is scale invariant so that a simple dimensional analysis can be applied. Throughout the paper we refer to some theories as a ``conformal field theory'', although we just need the property that it is scale invariant. While we are aware of the fact that this name is not always entirely appropriate, we do so in order to keep the analogy with the dual field theory in the AdS/CFT correspondence.} This conformal field theory is 
coupled to  fermions described by the spinor field $\chi$ and its conjugate $\bar{\chi}$.
In this case, the total system is modeled by the following action\footnote{In our notation, the space-time arguments of the fields in the (3+1)-dimensional Lagrangian density are always suppressed if the fields are evaluated at the same spacetime point $x^{\mu}$. Spinor indices are also suppressed and a sum over them is always implied.}
\begin{equation}\label{eq:critfermions}
S = S_{\text{cft}}[\Phi^*,\Phi] +  S_{0}[\bar{\chi},\chi] + i g \int d^4 x \,\bar{\chi}\chi\left(\Phi  + \Phi^*\right).
\end{equation}
The bosonic fluctuations are described by $\Phi$, an operator in the conformal field theory described by the action $S_{\text{cft}}$, whose expectation value is proportional to the order parameter. The noninteracting fermionic action is denoted by $S_{0}$ and $g$ is the coupling constant between the fermion and the conformal field theory.

When the bosonic excitations are free or weakly coupled, the full theory can be described elegantly using perturbation theory.
However, it is more challenging to study $S_{\text{cft}}$ in the case of a strongly coupled system. An example of a strongly coupled system with a continuous phase transition is a system of cold fermionic atoms at unitarity \cite{Zwerger12,Gubbels13}. At unitarity the scattering length parameterizing the interaction strength between the atoms becomes infinite, and there is no small parameter in which a perturbation expansion can be made.\footnote{This regime can however be accessed with renormalization-group techniques \cite{Gubbels08}.} For a positive chemical potential $\mu$, this system exhibits a continuous phase transition from a normal to a superfluid state at a nonzero temperature. The critical temperature can be represented as a line in the $(T,\mu)$-plane, which is suppressed as $\mu$ decreases and ends in a quantum critical point at zero temperature \cite{sachdev97}, see the right panel of Fig.~\ref{fig:phasecrit}.

A possible approach to obtain properties of a conformal field theory in the strongly coupled regime is derived from the holographic duality. Application of holographic methods to condensed-matter physics has been dubbed Anti-de Sitter/Condensed-Matter Theory (AdS/CMT) correspondence \cite{Herzog09,Hartnoll09,McGreevy10}.
Essentially, it boils down to the fact that correlation functions of operators in strongly coupled conformal field theories are provided by classical computations in a gravity dual of one dimension higher.
According to the discussion above, these correlation  functions can then be used to study the effective behavior of fermions in the vicinity of a quantum critical point.

With this application in mind, we recently described a model for interacting Dirac semimetals in the AdS/CMT set-up \cite{ARPES12,Weyl13}. This so-called dynamical-source model is constructed analogously to the theory in Eq.~(\ref{eq:critfermions}) and is similar to the semiholographic approach introduced in Ref.~\cite{Faulkner11}. It contains elementary Dirac fermions living in (3+1)-dimensional Minkowski space that are coupled to a conformal field theory playing the role of the critical system. Here, elementary means that the fermionic creation- and annihilation operators satisfy the canonical equal-time anticommutation relations given by
\begin{equation*}
\left\{\hat{\chi}(\vec{x},x^0) ,\hat{\chi}^{\dagger}(\vec{x}',x^0)\right\}= \delta^3\big(\vec{x}-\vec{x}'\big).
\end{equation*}
The conformal field theory they are coupled to, is in fact the dual (``boundary'') field theory of classical Einstein gravity in a (``bulk'') (4+1)-dimensional asymptotically Anti-de Sitter background with a planar Schwarzschild black hole. The main result is the retarded propagator of elementary Dirac fermions which contains a free part and a nontrivial self-energy. This self-energy comes about from adding probe Dirac fermions to the theory and integrating out the holographic dual conformal field theory part of the theory. This is why a boundary interpretation of the dynamical-source model is closely related to the system described in Eq.~(\ref{eq:critfermions}). The boundary Dirac fermions are made dynamical by an additional boundary term added to the fermionic action, which is an irrelevant perturbation to the conformal field theory, and its role is to provide the correct UV dynamics. As a consequence, the resulting retarded Green's function of the elementary fermionic operators defined by
 \begin{equation*}
 G_R(x-x') =
 -i \theta\left(x^0-x'^0\right)\Big\langle \Big\{\hat{\chi}(x), \hat{\chi}^{\dagger}(x') \Big\}\Big\rangle,
 \end{equation*}
 satisfies the zeroth-order frequency sum rule, i.e.,
\begin{equation}\label{eq:sumrule}
\int_{-\infty}^{\infty} \frac{d\omega}{c} \,\text{Im} \,G_R(\vec{k},\omega^+) = -\pi.
\end{equation}
This is important in the light of condensed-matter applications, where this sum rule is required for the determination of experimentally accessible correlators. So these holographic correlation functions can be directly compared to experimental data, e.g. from photoemission experiments or from radio-frequency spectroscopy for fermionic atoms at unitarity.

The dynamical-source model is an example from a bigger class of models called holographic bottom-up models. This is a phenomenological approach to the holographic duality, which is an alternative to the so-called top-down approach.
 In top-down holography, the starting point is string theory in ten dimensions with a set of $D$-branes. The benchmark example is the duality between Type-IIB string theory in AdS$_5 \times$S$_5$, and $\mathcal{N}=4$ Super-Yang-Mills theory in (3+1)-dimensional Minkowski space. By taking certain limits, consistent truncations to a lower-dimensional gravity theory can be made. If the string theory has a dual conformal field theory, the truncated field theory on the boundary is guaranteed to exist. However, the dimensional reduction usually leads to many technical difficulties, for instance, a large amount of degrees of freedom that have to be kept track of.
It is usually not clear which of these degrees of freedom are important for the behavior of the boundary field theory. Therefore, practical computations are often not feasible in such a set-up. See, however, Ref.~\cite{erdmenger13} for a recent example of a top-down approach to a condensed-matter problem.

 In contrast, in bottom-up models, we start off on the gravity side with classical Einstein gravity, such as an (asymptotically) Anti-de Sitter space-time. A few degrees of freedom are then put in by hand, for instance, a scalar field, the aforementioned Dirac field, or a Maxwell field. These represent the desired features of the strongly coupled field theory of interest, and it is a working assumption that this field theory exists. It is an effective low-energy theory, and there is no guarantee that these gravity theories can be embedded in a higher-dimensional string theory, where the duality was originally conjectured. It is for instance not clear that the Dirac fermions come from a certain limit of a string theory. Bottom-up holography is therefore a phenomenological approach, usually the subject of trial and error until successful, plausible or consistent results justify this approach. The upshot is that computations are relatively simple, without the necessity to deal with the many technicalities coming from string theory, but still manage to capture the universal and effective properties of many systems. In fact, this simplicity is particularly appealing for condensed-matter physics, where a description in terms of complicated supersymmetric non-Abelian gauge theories seems a bit of an overkill. The relevance of this phenomenological approach for the study of the low-energy physics of condensed-matter systems, was first stressed by the authors of Ref.~\cite{kiritsis10}.

Most AdS/CMT models are bottom-up models. So for applications in condensed matter, bottom-up holography in the form of the AdS/CMT approach may be the preferred way to go, compared to the top-down approach. Nevertheless, even in the bottom-up case, the question remains whether the dual boundary theory in fact provides an adequate description of a specific condensed-matter system.
 The hope is that the dual strongly coupled quantum field theory bears sufficient resemblance to realistic quantum critical points, such as the ones in the examples above, and understanding this point is an ongoing challenge. An important approach to comprehend bottom-up holography from the field-theory side, is to investigate what are the minimal requirements for conformal field theories to have a gravity dual. Examples are Refs.~\cite{Liu98,Balasubramanian99,DHoker99}, and more recently, Refs.~\cite{Heemskerk09,Fitzpatrick11,Papadodimas12}, and references therein. These works are aimed at a better understanding of the so-called conformal bootstrap constraints in general conformal field theories that have a large number of degrees of freedom. This information can be used to analyze the bulk effective action, by comparing the structure of correlation functions computed directly in the conformal field theory and in bottom-up holography using Witten diagrams. Nevertheless, the difficulty remains, that in the bottom-up approach, the AdS/CMT correspondence just provides the correlation functions of the effective excitations in the boundary, without providing information about their microscopic nature. So the microscopic Hamiltonian of the actual boundary system remains hidden.
 That problem is the topic of this work. We investigate possible microscopic mechanisms in which AdS/CMT results at zero charge density, such as the fermionic self-energy in the dynamical-source model can come about. To this end, we construct field theories of fermions coupled to a (strongly interacting) conformal field theory which mimic the bottom-up dynamical-source model mentioned above, as well as the theory in Eq.~(\ref{eq:critfermions}). The precise form of these simple field theories is motivated by making analogies to known holographic dualities. By making certain choices, it is possible to reproduce the result for the fermionic self-energy and electrical conductivity obtained in the dynamical-source model and elsewhere in the literature.
 In this manner we aim at a better understanding of how holographic bottom-up results can originate from condensed-matter field theories. Our approach is different from that in Refs.~\cite{Liu98,Balasubramanian99,DHoker99,Heemskerk09,Fitzpatrick11,Papadodimas12} in the sense that we have in mind to find a possible microscopic theory of elementary fields of the conformal field theory, and only indirectly consider the structure of the conformal field theory operators and their correlators. Hopefully this paves the way for a more established bottom-up holographic dictionary, where the relation between microscopic field theories suitable for condensed-matter, and holographic bottom-up model building becomes more precise. This might enhance the potential applicability of holographic results e.g. in condensed-matter theory and experiment.

  The content of this paper is as follows. In section \ref{sec:lit}, we first review the results of a different bottom-up computation, namely of the electrical conductivity of a strongly coupled conformal field theory \cite{Kovtun08}. Next, we review the dynamical-source model for the electrical conductivity of Dirac fermions coupled to a strongly interacting conformal field theory. In section \ref{sec:pfmodel} we depart from holography and construct a purely field-theoretic model. This is referred to as the probe-fermion model, because it contains a conformal field theory coupled to a fermion in the spirit of Eq.~(\ref{eq:critfermions}). In addition, it contains a certain large-$N$ limit and a suitably defined strong-coupling limit analogous to what happens on the field-theory side of the holographic duality. For both literature results in section \ref{sec:lit}, we show that this purely field-theoretic model reproduces the holographic results, thus providing a possible microscopic interpretation.
  Finally, in section \ref{sec:fock} we realise that this possible interpretation has some disadvantages, and construct an alternative model that reproduces the AdS/CMT results. This last, so-called Fock, model then offers the interpretation of the dynamical-source model as a phenomenological way to incorporate finite-coupling corrections in the bulk setup. We end with a conclusion in section \ref{sec:discussion}.
  There are also a number of appendices which contain more explanation and details of the calculations. In particular, our notation and conventions can be found in appendix \ref{app:conventions}.

\section{Previous results}\label{sec:lit}

We start by briefly reviewing two literature results from the AdS/CMT correspondence. Firstly, the authors of Ref.~\cite{Kovtun08} computed the electrical conductivity of the strongly coupled boundary conformal field theory with a gravity dual. Secondly, in Ref.~\cite{ARPES12} results were presented for the self-energy of Dirac fermions coupled to a conformal field theory with the same gravity dual, in the dynamical-source model. With this fermionic self-energy, the contribution to the electrical conductivity of the dynamical-source fermions was also computed \cite{Dirac14}.
In all three cases, the bottom-up gravity dual is classical Einstein gravity with a negative cosmological constant, that has as a solution an asymptotically Anti-de Sitter space-time with a planar Schwarzschild black hole. The corresponding line element is in $4+1$ dimensions given by
\begin{equation}\label{eq:aadsmetric}
ds^2= - \frac{V^2(r) r^2}{\ell^2} c^2 dt^2 + \frac{\ell^2}{r^2 V^2(r)}dr^2 +
\frac{r^2}{\ell^2} d\vec{x}^2,
\end{equation}
where $\ell$ is the characteristic length scale of the asymptotically AdS spacetime. On top of this fixed background, extra fields are added, which are specified below.

\subsection{Electrical conductivity of the boundary field theory}
For the electrical conductivity of the boundary field theory, a Maxwell field is added to the bulk system. Its action is in 4+1 dimensions and in natural units given by
\begin{equation*}
S_{\text{em}} = - \frac{1}{4g_{5}^2 \ell}\int d^{5}x \sqrt{-g} F_{\mu\nu} F^{\mu\nu},
\end{equation*}
where $g_5^2$ is the 5-dimensional gauge coupling constant, $g$ is the determinant of the metric corresponding to the line element in Eq.~(\ref{eq:aadsmetric}), and $F$ is the electromagnetic field or Faraday tensor. Compared to Ref.~\cite{Kovtun08} there is an additional factor $\ell$ so that $g_5^2$ is dimensionless here. The spatial components of the gauge field are fluctuating on top of the background in Eq.~(\ref{eq:aadsmetric}), and the linearized equation of motion for the gauge field in the curved background is considered. According to the holographic dictionary, the local U(1) symmetry becomes a global U(1) symmetry in the boundary field theory. The boundary value of the gauge field couples to the boundary U(1) symmetry current, and from the bulk gauge-field fluctuations the retarded two-point function of the boundary current is derived.

The boundary field is effectively charged by a rescaling of the global boundary current $J_{\mu} \rightarrow e g_5 J_{\mu}$ so that it becomes the canonically normalized charge current. The conductivity is subsequently obtained via the Kubo formula and this results in 3+1 boundary dimensions in
\begin{equation}\label{eq:kovtunSI}
\sigma=\frac{e^2k_B T}{\pi  \hbar^2 c},
\end{equation}
 where $T$ is the temperature. In Eq.~(\ref{eq:kovtunSI}) we have reexpressed the result from Ref.~\cite{Kovtun08} in SI units. For our purposes the most important feature of this result is that the conductivity is linear in temperature. This is to be expected from dimensional analysis in the boundary conformal field theory.
 Note that the conductivity
in Eq.~(\ref{eq:kovtunSI}) does not depend on
$\ell$, which is a
consequence of the scale
invariance of the system.

\subsection{Semiholographic fermionic self-energy and fermionic conductivity}
In the computation of the self-energy of Dirac fermions coupled to the boundary field theory in the dynamical-source model \cite{ARPES12,Weyl13,Dirac14}, two species of uncoupled probe Dirac fields $\Psi^{(i)}$ with $i=1,2$ are added to the bulk instead of a Maxwell field. Special boundary conditions are used, in particular, next to the usual Dirichlet boundary conditions, an IR irrelevant kinetic boundary term for the Dirac fermions is taken into account. These boundary terms are imposed at a UV slice $r=r_0$, which is taken to the boundary at the end of the computation. The total action is
\begin{equation*}\begin{aligned}
S& =-i g_f  \sum_{i=1}^2\int d^5 x  \sqrt{-g}\bar{\Psi}^{(i)} \left(\frac{1}{2}\Gamma^{a}e_a^{\;\;\mu}\overleftrightarrow{D}_{\mu}- M_i\right)\Psi^{(i)} \\
& - \int d^4 x  \sqrt{-g} \bigg[ i g_f\sqrt{g^{rr}}\left(\bar{\Psi}_R^{(1)} \Psi^{(1)}_L - \bar{\Psi}^{(2)}_L \Psi_R^{(2)}\right)
 +Z \bar{\Psi} \slashed{D}_4 \Psi \bigg]\bigg|_{r=r_0}.
\end{aligned}\end{equation*}
 Here,
 $\Psi^{(i)}_{R,L}$ are the chiral components of $\Psi^{(i)}$ with respect to the chiral Dirac matrix of the boundary, and $\Psi = \Psi^{(1)}_R + \Psi^{(2)}_L$ is the boundary Dirac spinor. Furthermore, $g^{\mu\nu}$ and $e_a^{\;\mu}$ are respectively the components of the asymptotically Anti-de Sitter metric introduced in Eq.~(\ref{eq:aadsmetric}) and the corresponding vielbeins, $D_{\mu}$ is the usual covariant derivative containing the spin connection, $\slashed{D}_4=\sum_{a\neq r}\Gamma^{a}e_{a}^{\;\;\mu} i \dau_{\mu}$ is the boundary kinetic operator where $\Gamma^a$ are the bulk Dirac matrices, $g_f$ is a dimensionless coupling constant, and $Z$ is a dimensionful wavefunction-renormalization constant. Finally, $M_1=-M_2=M$, are the dimensionless bulk Dirac masses, which are in the interval $-1/2 < M < 1/2$. This number $M$ becomes a model parameter in the boundary field theory.

  Again, the equation of motion is solved in the curved background and the action is evaluated on shell.
  This implies that the boundary conformal field theory is integrated out, and it takes on the role of an effective self-energy for the boundary Dirac spinor $\Psi$. This self-energy can be written as the solution of the Dirac equation in the curved background. As a result of the Gaussian integration, half of the chiral components of both $\Psi^{(i)}$ are eliminated, namely $\Psi^{(1)}_L$ and $\Psi^{(2)}_R$. So the chiral component of each bulk fermion species that is left, provides one of the chiral components of $\Psi$. Next, a field rescaling and the limit $r_0\rightarrow\infty$ are carried out in a specific manner that keeps the boundary action finite. Together with the kinetic boundary term, the result is the retarded Green's function of Dirac fermions in 3+1 dimensions, with the retarded self-energy coming from the interactions with the strongly coupled conformal field theory. For zero temperature, the result is\footnote{Note that in order to make the self-energy agree with our present conventions, we have, compared to Ref.~\cite{Dirac14}, placed an additional minus sign in front of the action, and removed a factor $c\gamma^0$ from the inverse Green's function to arrive at Eqs.~(\ref{eq:Gretds}) and (\ref{eq:selfv0}). See also appendix \ref{app:conventions}. Furthermore, the dimensionless number $C_M$ in Eq.~(\ref{eq:dscond}) is equal to $\big[2^{2M}\Gamma(\frac{1}{2}~+~M)/\Gamma(\frac{1}{2}~-~M)\big]^2 S^{\text{IR}}_M$ in the notation of Ref.~\cite{Dirac14}.}
  \begin{equation}\label{eq:Gretds}
  G^{-1}(k) = \slashed{k} - \Sigma(k),
  \end{equation}
  with the self-energy
\begin{equation}\label{eq:selfv0}
\Sigma(k)= - \frac{\lambda}{2^{2M}c} \frac{\Gamma(\frac{1}{2}-M)}{\Gamma(\frac{1}{2}+M)} \slashed{k}k^{2M-1},
\end{equation}
 where $ck^{0} = \omega+i0$ and the dimensionful quantity $\lambda=c \ell^{2M} (r_0/\ell)^{2-2M}g_f/Z \geq 0$ is the square of the coupling between the conformal field theory and the fermion. The  anomalous dimension of the self-energy quantifies the difference from the linear relativistic scaling. So we see that $2M-1$ is related to the anomalous dimension of the self-energy.

Next, the dynamical-source fermion can be minimally coupled to a background electric field, and thus the contribution of the dynamical-source fermion to the electrical conductivity can be computed. In Ref.~\cite{Dirac14} this computation was carried out
 and for $\omega$ much smaller than both frequency scales present, namely, $\omega \ll k_B T/\hbar$ and $\omega \ll (\lambda/c^{2M})^{1/(1-2M)}$, the result was
\begin{equation}\label{eq:dscond}
\sigma_{\chi}
 = \frac{e^2 c^2}{12 \pi \hbar \lambda^2} \left(\frac{k_B T}{\hbar c}\right)^{3-4M} C_M.
\end{equation}
Here, $C_M$ is a dimensionless function of $M$. In contrast to the pure conformal field theory contribution from Eq.~(\ref{eq:kovtunSI}), this contribution scales as temperature to the power $3-4M$. So, the conductivity obtains a linear scaling in temperature only in the limit $M\rightarrow 1/2$. The point $M=1/2$ thus corresponds to a case with no anomalous dimension, as is also visible in Eq.~(\ref{eq:selfv0}). In fact, a different calculation has to be carried out in this case, which results in logarithmic corrections to a linear power-law scaling.

\section{Probe-fermion model}\label{sec:pfmodel}

In an attempt to interpret the above results, we now construct a simple field theory, the probe-fermion model.
The total action of this model is\footnote{For us, the inspiration to write down an action of this specific form came from the structure of supersymmetric Yang-Mills theory.
For the readers without a background in supersymmetry, we have included appendix \ref{app:sym}, in which we review some properties of supersymmetric Yang-Mills theory that may motivate the action in Eq.~(\ref{eq:Stotpfmodel}).}
\begin{equation}\label{eq:Stotpfmodel}
S_\text{tot}[\varphi,\psib,\psi;A,\bar{\chi},\chi]= S_0[\varphi]+S_0[\psib,\psi;A] +S_{g_2}[\varphi,\psib,\psi] + S_0[\bar{\chi},\chi;A] + S_{g_1}[\varphi,\psib,\psi;\bar{\chi},\chi].
\end{equation}
The notation indicates that the action is a functional of the sets of fields $\varphi_i$, and $\psi_i$ and $\psib_i$, and of the sources $A$, and $\chi$ and $\bar{\chi}$. These ingredients are specified in the following.

Firstly, the fields $\psi_i$, $\psib_i$ and $\varphi_i$
together are referred to as the ``conformal field theory''. As discussed in the introduction, this name is not completely appropriate. It refers here to the scale-invariant theory that takes over the role of the field theory dual in the actual holographic model. The theory has a number of copies of fields $i=1,...,N$. Furthermore, the coupling constants $g_2$ and $g_1$ introduced below are normalized in a certain way, by appropriate factors of $1/\sqrt{N}$. This normalization makes sure that a certain class of diagrams dominates in the large-$N$ limit.
The $\varphi_i$ are real scalars whose contribution is given by the action
\begin{equation*}
S_0[\varphi]=-\frac{\hbar}{2}\sum_{i=1}^{N} \int d^4 x \,\varphi_i  \left(-i\dau\right)^{2-\eta}\varphi_i.
\end{equation*}
The propagator of the $\varphi_i$ has an anomalous dimension $\eta$.
Next, we have Dirac fermions $\psi_i$ and their conjugates $\psib_i=\psi_i^{\dagger}\gamma^0$. Writing $\slashed{\dau} = \gamma^{\mu}\dau_{\mu}$ and $\slashed{A} = \gamma^{\mu}A_{\mu}$, they are described by the action
\begin{equation}\label{eq:Spsi}
 S_0[\psib,\psi;A]= - i \hbar  \sum_{i=1}^{N}\int d^4x \,\bar{\psi}_{i} \Big(\slashed{\dau}-\frac{ie}{\hbar}\slashed{A}\Big)\psi_i.
\end{equation}
The components of the spinors $\psi_{i}$ and $\psib_i$ are Grassmann fields.
The Dirac fermions are coupled to the real scalars through a local four-point interaction with coupling constant $g_2/2N$, described by the term
\begin{equation}\label{eq:Sg2}
S_{g_2}[\varphi,\psib,\psi]=\frac{i\hbar g_2}{2N} \sum_{i=1}^{N} \sum_{i'=1}^{N}\int d^4 x \, \varphi_i\,\bar{\psi}_i\,\psi_{i'}\,\varphi_{i'}.
\end{equation}

There are also two ``external'' fields, i.e., not in the conformal field theory, which we motivate now. Firstly, in section \ref{sec:elecft}, we compute the electrical conductivity of the field theory. To this end, the
Dirac fermions $\psi_i$ in Eq.~(\ref{eq:Spsi}) are minimally coupled with coupling $e$ to a nondynamical U(1) gauge field $A_{\mu}$. The scalar field $\varphi$ is real and is not minimally coupled.
Secondly, in section \ref{sec:selfft}, the fermionic self-energy is computed. To calculate it in the probe-fermion model, we add a single Dirac fermionic source $\chi$ to the theory with noninteracting action
\begin{equation}\label{eq:Schi}
S_0[\bar{\chi},\chi;A]=- i \hbar  \int d^4 x \, \bar{\chi} \Big(\slashed{\dau}-\frac{ie}{\hbar}\slashed{A}\Big)\chi.
\end{equation}
Analogously to the computation in the dynamical-source model discussed in section \ref{sec:lit} this is the external Dirac fermion that is coupled to a conformal field theory and obtains a nontrivial self-energy when the field theory is integrated out. In other words, this $\chi$ fermion is the probe fermion that mimics the dynamical source. For generality, it is also minimally coupled to the U(1) gauge field.
There is only a single species of this fermion because it is not part of the conformal field theory in which there is a large-$N$ limit. Instead, the external source $\chi$ is coupled to the field theory through a local three-point interaction with coupling $g_1/\sqrt{N}$, described by the term
\begin{equation*}
S_{g_1}[\varphi,\psib,\psi;\bar{\chi},\chi]=\frac{i\hbar g_1}{\sqrt{N}}\sum_{i=1}^{N}\int d^4 x  \bigg(\varphi_i\,\bar{\chi}\,\psi_i+ \bar{\psi}_i\,\chi\,\varphi_i\bigg).
\end{equation*}
Thus, we aim at an interpretation of the dynamical-source model, in which the dynamical-source fermion is a probe fermion coupled to a conformal field theory.

The real-time partition function is now
\begin{equation*}\begin{aligned}
&Z[A,\bar{\chi},\chi]=\int d[\psib] d[\psi]\int d[\varphi] \,\exp\bigg(\frac{i}{\hbar}S_{\text{tot}}[\varphi,\psib,\psi;A,\bar{\chi},\chi]\bigg).
\end{aligned}\end{equation*}
Our notation for the path-integral measure is $d[\text{field}] = \prod_{i=1}^{N} d[\text{field}_i]$.
To perform the large-$N$ limit we perform a Hubbard-Stratonovich transformation to a collective spinor field $\pi$ that decouples both interactions. To this end, the partition function is multiplied by
\begin{equation}\label{eq:HSpi}\begin{aligned}
1 =  \int d[\bar{\pi}]d[\pi]\exp\bigg[&\frac{i}{\hbar}\int d^4x \bigg(\bar{\pi}-\frac{\hbar g_2}{2N} \sum_i \bar{\psi}_i \varphi_i - \frac{\hbar g_1}{\sqrt{N}} \bar{\chi}\bigg)\bigg(\frac{2N}{i\hbar g_2}\bigg)\\
&\times\bigg(\pi - \frac{\hbar g_2}{2N}\sum_{i'} \psi_{i'} \varphi_{i'} - \frac{\hbar g_1}{\sqrt{N}}\chi\bigg)\bigg].
\end{aligned}\end{equation}
Then, the partition function becomes
\begin{equation}\label{eq:Z1}\begin{aligned}
&Z[A,\bar{\chi},\chi]=\int d[\psib]d[\psi]\int d[\varphi]\int d[\bar{\pi}] d[\pi] \,\exp\bigg(\frac{i}{\hbar}S^{\text{HS}}_{\text{tot}}[\varphi,\psib,\psi,\bar{\pi},\pi;A,\bar{\chi},\chi]\bigg),
\end{aligned}\end{equation}
with
\begin{equation}\label{eq:Seff1}\begin{aligned}
&S^{\text{HS}}_{\text{tot}}[\varphi,\psib,\psi,\bar{\pi},\pi;A,\bar{\chi},\chi] = S_0[\varphi]+S_0[\psib,\psi;A] 
+ \int d^4 x\bigg[\bar{\pi}\Big(\frac{2N}{i\hbar g_2}\Big)\pi\\
 &+ \bar{\chi}\left[-i \hbar  \Big(\slashed{\dau}-\frac{ie}{\hbar} \slashed{A}\Big)-\frac{2i\hbar g_1^2}{g_2}\right] \chi +\frac{2i \sqrt{N} g_1}{g_2}\big(\bar{\chi}\pi + \bar{\pi} \chi\big) +  i\sum_i\Big(\bar{\pi}  \varphi_i \psi_i + \bar{\psi}_i \varphi_i \pi\Big) \bigg].
\end{aligned}\end{equation}

Starting from Eq.~(\ref{eq:Seff1}) our approach is the following. Now that the interactions are decoupled, the only $N$-dependence of the effective actions of the $\varphi_i$ and $\psi_i$ fields is an overall species sum. This just leads to $N$ copies of the corresponding single-species partition functions. Furthermore, these actions are Gaussian, so we can integrate out the $\varphi_i$ and $\psi_i$ fields exactly. This mimics the integrating out of the conformal field theory in the semiholographic dynamical-source model. After this, we work in the random-phase approximation, which becomes exact when the large-$N$ limit is taken. Then the collective field can be integrated out. The result is the generating functional in Eq.~(\ref{eq:Z3}) below, from which the conductivity and the self-energy of the probe fermion can be obtained.
We will now derive Eq.~(\ref{eq:Z3}) in several steps.

Before we integrate out anything, it is convenient to introduce some new notation. We first define the inverse noninteracting
Green's function corresponding to a single species of the scalar fields $\varphi_i$ as
\begin{equation}\label{eq:Gphix}
G^{-1}_{\varphi}(x,x') = -\big(-i \dau\big)^{2-\eta} \delta^4(x-x').
\end{equation}
The inverse noninteracting Green's function for a single species of the fermion fields $\psi_i$ is a functional of $A_{\mu}$, since
\begin{equation}\label{eq:Gpsix}
G^{-1}_{\psi}[A](x,x') = -i \left(\slashed{\dau}-\frac{ie}{\hbar}\slashed{A}\right)\delta^4(x-x').
\end{equation}
It is also possible to write the Green's function in their corresponding momentum-space representation. However, it is more elegant to go to a basis-independent description, where the Green's functions matrices are written without indices. We also introduce a matrix inner product, which is in the coordinate representation defined as:
\begin{equation*}
\big(\phi \big| G_{\phi}^{-1} \big| \phi \big)\equiv \int d^4 x \int d^4 x'\;\bar{\phi}(x)\,G^{-1}_{\phi}(x,x')\, \phi(x'),
\end{equation*}
where $\phi$ denotes any of the fields with a suitably defined conjugate $\bar{\phi}$, which should be clear from the context. As before, a sum over spinor indices is always assumed if appropriate. Furthermore, all Green's functions are diagonal in species indices.
With this definition, the quadratic part of the actions can be written as a basis-independent inner product. For example, the noninteracting fermion action from Eq.~(\ref{eq:Spsi}) becomes
\begin{equation*}
S_0[\psib,\psi;A]=  \hbar \sum_{i=1}^N \big(\psi_i \big| G_{\psi}^{-1}[A] \big| \psi_i\big).
\end{equation*}
Now we integrate out the $\psi_i$ fermions from Eq.~(\ref{eq:Seff1}), after which the effective action for the scalar fields is written in basis-independent notation as
\begin{equation*}
S^{\text{eff}}[\varphi,\bar{\pi},\pi;A]=\frac{\hbar}{2}\sum_{i=1}^N \big(\varphi_i \big| G_{\varphi}^{-1}-\Sigma_{\varphi}[\bar{\pi},\pi,A]\big| \varphi_i\big),
\end{equation*}
The scalars have an effective self-energy, which is in the coordinate representation given by
\begin{equation*}
\Sigma_{\varphi}[\bar{\pi},\pi,A](x,x')= -\frac{2}{\hbar^2}\bar{\pi}(x)\, G_{\psi}[A](x,x') \, \pi(x').
\end{equation*} Now it is straightforward to integrate out the scalar fields, which yields
\begin{equation}\label{eq:gluonint}
\int d[\varphi] \exp\bigg[i \sum_{i=1}^N \big(\varphi_i \big| \frac{1}{2}G_{\varphi}^{-1}-\Sigma_{\varphi}[\bar{\pi},\pi,A]\big| \varphi_i\big)\bigg]= \exp\bigg[-\frac{N}{2} \Tr\ln\Big(- G_{\varphi}^{-1} + \Sigma_{\varphi}[\bar{\pi},\pi,A]\Big)\bigg].
\end{equation}

At this point we perform a fluctuation expansion of $\pi$ around its expectation value $\langle \pi\rangle$. Here we expand around $\langle \pi\rangle=0$ which is always a solution of the saddle-point equation.
To obtain a nonzero result we thus have to go one order beyond the mean-field approximation, which is the random-phase approximation mentioned earlier. Because of the overall factor $N$ in the exponent above, the random-phase approximation becomes exact in the large-$N$ limit.
Taking the large-$N$ limit, we can therefore work up to first order in the scalar self-energy and still obtain an exact result. Expanding the logarithm in Eq.~(\ref{eq:gluonint}) to this order yields
\begin{equation*}
\Tr\ln\Big(-G_{\varphi}^{-1} +\Sigma_{\varphi}\Big) = \Tr\ln\Big(-G_{\varphi}^{-1}\Big) +\Tr\ln\Big(1-G_{\varphi}\Sigma_{\varphi}\Big) \simeq \Tr\ln\Big(-G_{\varphi}^{-1}\Big)  - \Tr \,G_{\varphi}\Sigma_{\varphi}.
\end{equation*}
Here, $\Sigma_{\varphi}$ is still a functional of $\pi$ and $\bar{\pi}$, and this is used to obtain a self-energy term for the $\pi$ fluctuations. After these manipulations, the total partition function from Eq.~(\ref{eq:Z1}) becomes
\begin{equation}\label{eq:Z2}\begin{aligned}
&Z[A,\bar{\chi},\chi] = \exp\bigg[N \Tr\ln \Big(-G_{\psi}^{-1}[A]\Big)
-\frac{N}{2}\Tr\ln\Big(-G_{\varphi}^{-1}\Big)\bigg]\\
&\times \int d[\bar{\pi}] d[\pi] \,\exp\bigg[ i N \big(\pi\big|G_{\pi}^{-1}[A]\big|\pi \big)+ i\big(\chi\big| \mathcal{G}_{\chi}^{-1}[A]
\big|\chi\big) + \frac{i}{\hbar} \frac{2g_1\sqrt{N}}{g_2}\Big[i\big(\chi\big|\pi\big)+i\big(\pi\big|\chi\big)\Big]\bigg].
\end{aligned}\end{equation}
Here, we have used the notation
\begin{equation*}
 \big(\chi\big|\pi\big) = \int d^4 x \, \bar{\chi}(x)  \pi(x),
 \end{equation*}
 and the noninteracting Green's function of the probe fermion is given in coordinate space by\footnote{The symbol $G_{\chi}$ is reserved for the fully dressed probe-fermion Green's function.}
  \begin{equation}\label{eq:mathcalGchi}
 \mathcal{G}_{\chi}^{-1}[A](x,x') = \left[ -i \Big(\slashed{\dau} -\frac{ie}{\hbar} \slashed{A}\Big)-\frac{2ig_1^2}{g_2}
 \right]\delta^4(x-x').
\end{equation}
We have also defined the inverse collective-field propagator $G^{-1}_{\pi}$, which is a functional of $A_{\mu}$ through $G_{\psi}$ and is given in momentum space by
\begin{equation*}
G^{-1}_{\pi}[A](k) = \frac{2}{i\hbar^2 g_2}+\frac{i }{\hbar^2} \int \frac{d^4 q}{(2\pi)^4}\, G_{\psi}[A](k+q) \,G_{\varphi}(q).
\end{equation*}
Finally, we integrate out the collective-field fluctuations $\pi$, which reduces Eq.~(\ref{eq:Z2}) to the generating functional
\begin{equation}\label{eq:Z3}\begin{aligned}
Z[A,\bar{\chi},\chi]  &= \exp\bigg[N \Tr\ln \Big(-G_{\psi}^{-1}[A]\Big)
-\frac{N}{2}\Tr\ln\Big(-G_{\varphi}^{-1}\Big)+\Tr\ln\Big(-N G_{\pi}^{-1}[A]\Big)\bigg]\\
&\times\exp\bigg[i \big(\chi\big| \,\mathcal{G}_{\chi}^{-1}[A] +
\left(\frac{2 g_1 }{\hbar g_2}\right)^2\,G_{\pi}[A]\,\big|\chi\big) \bigg].
\end{aligned}\end{equation}
This generating functional is the starting point for computing the self-energy of the probe fermion and the electrical conductivity of the probe-fermion model.

\subsection{Self-energy of the probe fermion}\label{sec:selfft}

\begin{figure}
\vskip 10 pt
\centering
\includegraphics[width=.98\textwidth]{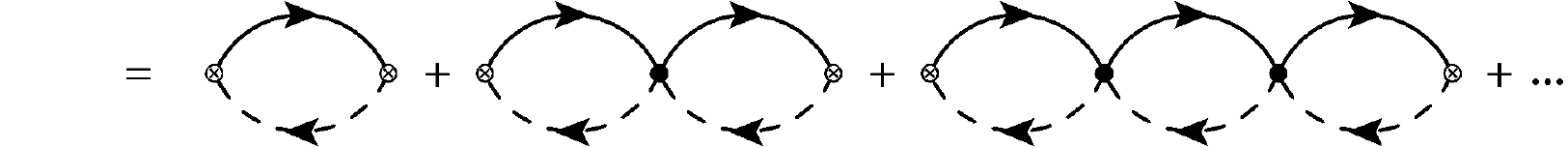}
\begin{picture}(0,0)
\put(-420,16){\Large{$\Sigma_{\chi}$}}
\end{picture}
\caption{Probe-fermion self-energy from Eq.~(\ref{eq:selfv1series}) in a perturbation series in powers of $g_2$, denoted by the filled cicles. Empty circles represent the three-point vertex $g_1$, where $\chi$ can attach. The solid line is the noninteracting Green's function of a single species of $\psi_i$ fermions, while the dashed line denotes the noninteracting scalar Green's function. Together, they form the bubble corresponding to the expression for $I(k)$ from Eq.~(\ref{eq:I}).}
\label{fig:probe_selfenergy_pert}
\end{figure}

For the self-energy of the probe fermion we consider Eq.~(\ref{eq:Z3}) and set the field $A_{\mu}$ to zero. Going to momentum space, the effective action for the probe fermion is identified as
\begin{equation}\label{eq:Schieff}
S^{\text{eff}}[\bar{\chi},\chi]=\hbar \int \frac{d^4 k}{(2\pi)^4}\, \bar{\chi}(k)\bigg[ \slashed{k} - \frac{2ig_1^2}{g_2}+ \frac{2ig_1^2}{g_2} \frac{1}{1-\frac{ g_2}{2} I(k)}\bigg]\chi(k),
\end{equation}
where
\begin{equation}\label{eq:I}
I(k) = \int \frac{d^4q}{(2\pi)^4}\, G_{\psi}(k+q) \,G_{\varphi}(q).
\end{equation}
From Eq.~(\ref{eq:Schieff}) we can read off the self-energy of $\chi$, which is
\begin{equation}\label{eq:selfv1}
\Sigma_{\chi}(k)= \frac{2ig_1^2}{g_2}  \left(1-\frac{1}{1-\frac{g_2 }{2}I(k)}\right)=\frac{-ig_1^2 I(k)}{1-\frac{g_2}{2} I(k)}.
\end{equation}
Note in particular that all factors of $N$ have dropped out from the self-energy. The result in Eq.~(\ref{eq:selfv1}) can be seen as a resummation of the bubble sum, where the bubble diagram corresponds to $I(k)$, the internal vertices come with a factor $g_2$ and the exterior vertices with $g_1$. Namely, when we expand in powers of $g_2$ it becomes a geometric series,
\begin{equation}\label{eq:selfv1series}\begin{aligned}
\Sigma_{\chi}(k)&\simeq -i g_1^2 I(k)\left(1+\frac{g_2}{2} I(k)+ \frac{g_2^2}{4} I^2(k)+...\right).
\end{aligned}\end{equation}
This series is depicted in Fig.~\ref{fig:probe_selfenergy_pert}.

The loop integral $I$ from Eq.~(\ref{eq:I}) is computed in appendix \ref{app:pfloop}.
At zero temperature, the result is given by
\begin{equation*}
I(k) =  \frac{i \slashed{k} \big(k^2\big)^{\frac{\eta}{2}}}{  (4\pi)^2} \,\frac{1}{\frac{\eta}{2}\big(1+\frac{\eta}{2}\big)\big(2+\frac{\eta}{2}\big)} \equiv i\slashed{k} \big(k^2\big)^{\frac{\eta}{2}} f(\eta),
\end{equation*}
where we defined the function $f$ as
\begin{equation*}
f(\eta)=\left[(4\pi)^2 \,\frac{\eta}{2}\left(1+\frac{\eta}{2}\right)\left(2+\frac{\eta}{2}\right)\right]^{-1}.
\end{equation*}
Inverting the matrix structure in Eq.~(\ref{eq:selfv1}) leads to two contributions to the self-energy,
\begin{equation}\label{eq:selfv2}
\Sigma_{\chi}(k)=-\frac{i g_1^2}{1+\left(\frac{g_2}{2}\right)^2(k^2)^{1+\eta}f^2(\eta)}\left[i \slashed{k} (k^2)^{\frac{\eta}{2}} f(\eta)-\frac{g_2}{2}(k^2)^{1+\eta}f(\eta)\unit\right].
\end{equation}

 We now define a strong-coupling limit as follows. The couplings $g_1$ and $g_2$ become infinite, but the ratio $g=g_1^2/g_2^2$ remains finite. Writing $g_1^2=g g_2^2$, we see that this particular limit amounts to ignoring the 1 in the denominators of Eq.~(\ref{eq:selfv2}). Then, the probe-fermion self-energy becomes
\begin{equation*}
\Sigma_{\chi}(k)=\frac{4 g}{f(\eta)} \slashed{k} \big(k^2\big)^{-\frac{\eta}{2}-1} + 2i g g_2\unit.
\end{equation*}
The second term is independent of momentum and is a formally infinite mass term, generated in the Hubbard-Stratonovich transformation. We can get rid of this term by adding a bare mass term for $\chi$ from the start, which precisely cancels this term in the strong-coupling limit. Then, the renormalized self-energy is just given by the first term, i.e.,
\begin{equation}\label{eq:selfv4}
\Sigma_{\chi}(k)=\frac{4 g}{f(\eta)} \slashed{k} \big(k^2\big)^{-\frac{\eta}{2}-1}.
\end{equation}
It is proportional to the product of $\slashed{k}$ and a power of $k^2$ related to the anomalous dimensions of the scalar fields.
This is precisely the form of the self-energy from Eq.~(\ref{eq:selfv0}) found in Ref.~\cite{ARPES12}, if we make the identifications
\begin{equation}\label{eq:pfdstranslate}\begin{aligned}
& \eta = -2M-1,\\
&g
 =- \frac{\lambda \,\Gamma\left(\frac{1}{2}-M\right)f(-2M-1)}{c\,4^{M+1}\Gamma\left(\frac{1}{2}+M\right) }.
\end{aligned}\end{equation}
The desired range $-1/2<M<1/2$ coincides with the interval $-2<\eta<0$.

We end with the remark that the case with no anomalous dimension, $\eta=0$, corresponds to $M=-1/2$, and not to $M=1/2$ as anticipated. Instead, the value $M=1/2$ corresponds to $\eta=-2$. Note that the holographic result from Eq.~(\ref{eq:selfv0}) is obtained in the alternative quantization. In the regular quantization, the self-energy from Eq.~(\ref{eq:selfv0}) would be replaced by its inverse. This is equivalent to Eq.~(\ref{eq:selfv0}) with the replacement of $M$ by $-M$. So in the regular quantization, $\eta=0$ indeed corresponds to $M=1/2$. Nonetheless, in the probe-fermion model result from Eq.~(\ref{eq:selfv4}), setting $\eta=0$ still does not make the self-energy proportional to $\slashed{k}$, which is the expected form without an anomalous dimension.

\subsection{Electrical conductivity}\label{sec:elecft}

In this section we consider the total conductivity in the probe-fermion model in linear-response theory. This is formally achieved by making use of the fact that the total electric current $J^{\mu}$ in the Lagrangian of the probe-fermion model couples to a nondynamical U(1) gauge field $A_{\mu}$. In practice, the total current consists of several contributions, which are defined through minimal coupling of the various charged fields to $A_{\mu}$, for instance as in Eqs.~(\ref{eq:Spsi}) and (\ref{eq:Schi}).

 In linear response, the conductivity tensor $\sigma^{ij}$ is defined as
\begin{equation}\label{eq:linresponse}
\Big\langle J^{i}(\kv,\omega)\Big\rangle = \sum_{j=1}^3 \sigma^{ij}(\kv,\omega) E^{j}(\kv,\omega),
\end{equation}
where $i,j = 1,2,3$, and we choose the temporal gauge where $A_0=0$. Furthermore, $E^{j} = i\omega A^{j}$ is a sufficiently small electric field. Because $J^{\mu}$ couples to $A_{\mu}$ in the functional integral, we have from  Eq.~(\ref{eq:linresponse}) that
\begin{equation*}
\sigma^{\mu \nu} \propto \frac{ \delta^2 \ln Z[A]}{\delta A_{\nu} \delta A_{\mu}}\bigg|_{A=0} \propto \Big\langle J^{\mu} J^{\nu} \Big\rangle.
\end{equation*}
So the generating functional Eq.~(\ref{eq:Z3}) is only needed up to second order in $A_{\mu}$. For the discussion in this section and in section \ref{sec:elecfock} it is sufficient to consider only $\delta^2 \ln Z/\delta A_{\mu}\delta A_{\nu}$. Obtaining the conductivity from this is straightforward and will not be discussed in detail here.
To consider the total contribution we start with the generating functional $Z[A,\bar{\chi},\chi]$ from Eq.~(\ref{eq:Z3}), and integrate out the fields $\bar{\chi}$ and $\chi$. The result is $Z[A]$ where all matter fields are integrated out. The $A$-dependent part is thus given by
\begin{equation}\label{eq:lnZ1}
\ln Z[A] = N \Tr \ln \Big(-G_{\psi}^{-1}[A]\Big) + \Tr \ln\Big(-N G_{\pi}^{-1}[A]\Big) + \Tr \ln\Big(- G^{-1}_{\chi}[A]\Big),
\end{equation}
where $G_{\chi}^{-1}$ is the full Green's function of the probe fermion, given by the quadratic part in $\chi$ and $\bar{\chi}$ in Eq.~(\ref{eq:Z3}),
\begin{equation*}
G^{-1}_{\chi}[A](x,x') = \mathcal{G}^{-1}_{\chi}[A](x,x') - \left(\frac{2g_1}{\hbar g_2}\right)^2  G_{\pi}[A](x,x').
\end{equation*}

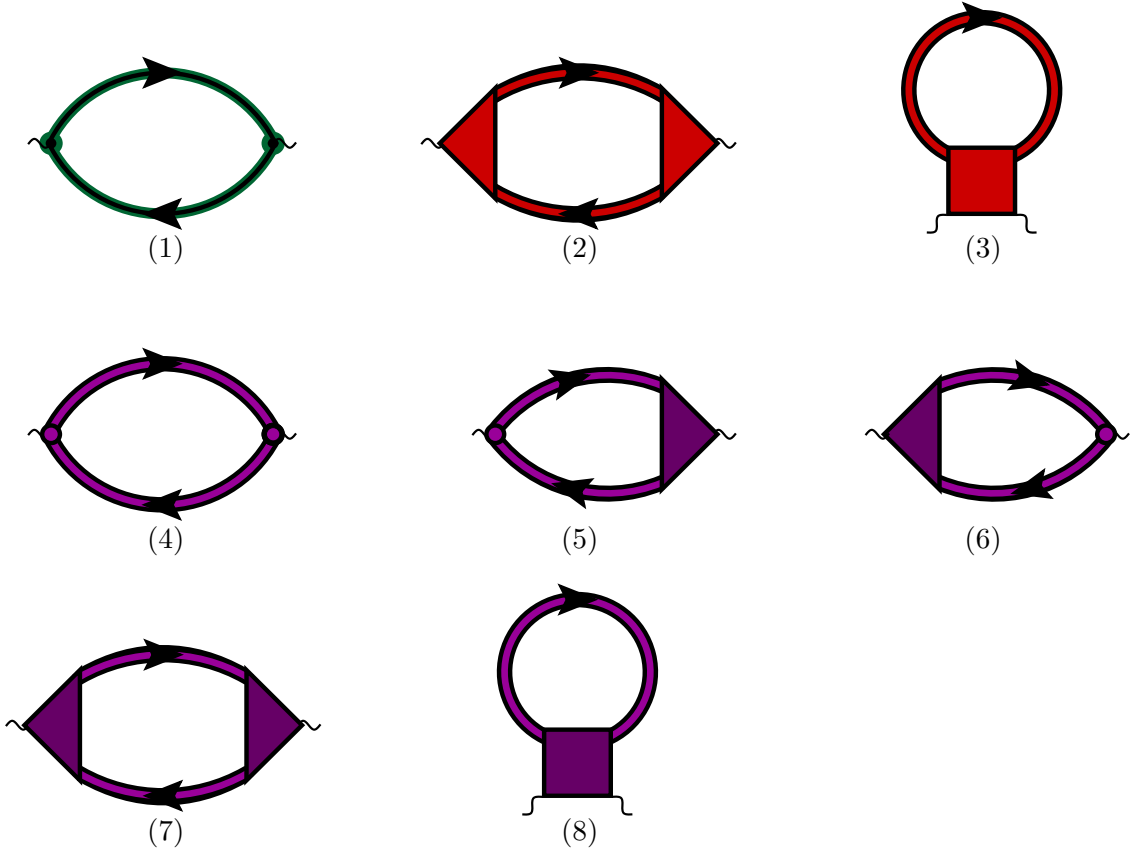
\begin{figure}[h!]
\vskip 10 pt
\centering
\includegraphics[width=.98\textwidth]{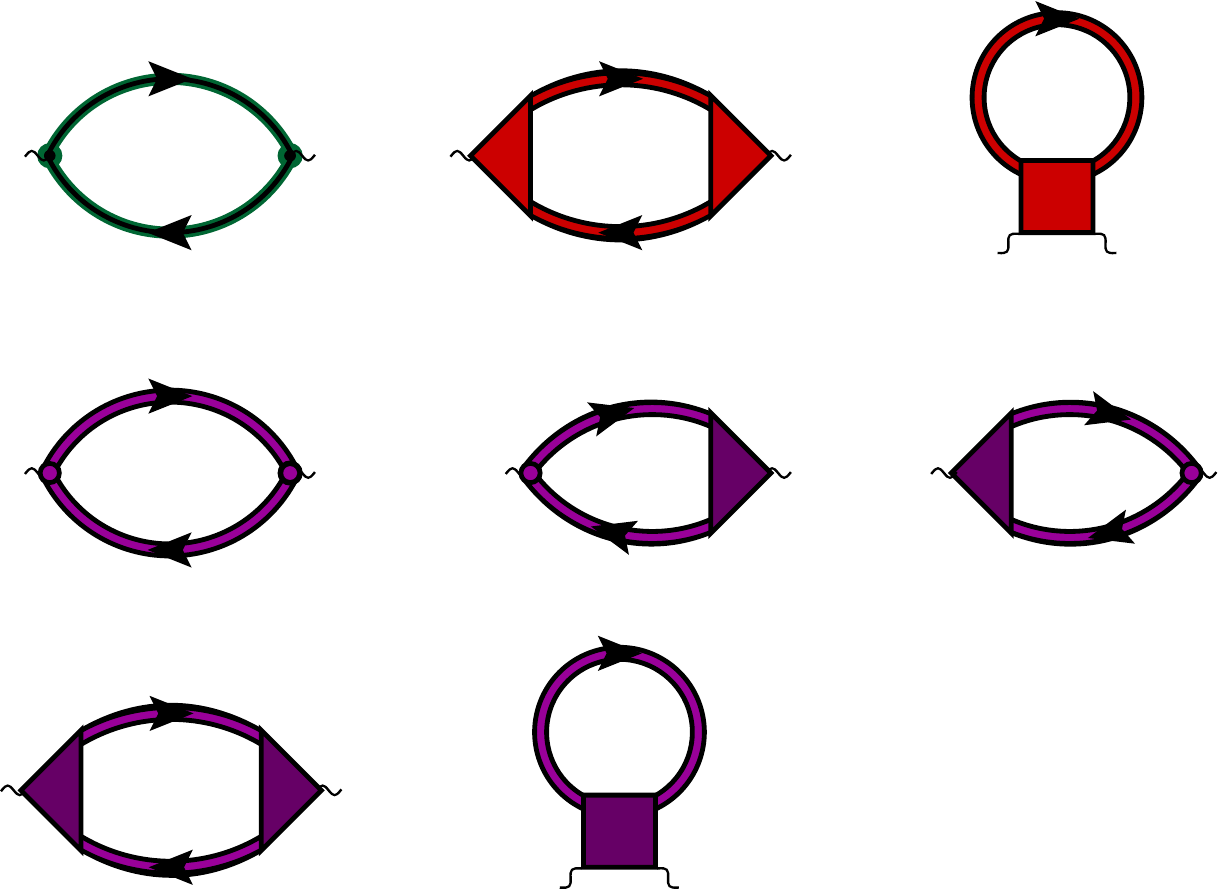}
\begin{picture}(0,0)
\put(-372,212){(1)}
\put(-217,212){(2)}
\put(-66,212){(3)}
\put(-372,102){(4)}
\put(-217,102){(5)}
\put(-66,102){(6)}
\put(-372,-8){(7)}
\put(-217,-8){(8)}
\end{picture}
\vskip 10pt
\caption{All diagrams corresponding to Eq.~(\ref{eq:dlnZ2pf}) in appendix \ref{app:lnZA} contributing to the total current-current correlation function in the probe-fermion model. The Feynman rules are as follows. Small wiggly lines denote the locations where external photons with vector indices $\mu$ and $\nu$ attach. The single solid (green) line is $G_{\psi}[0]$. The green vertex is $e\gamma^{\mu}/\hbar$. The double (red) line is $G_{\pi}[0]$. The red three-point vertex is $\delta \Sigma_{\pi}/\delta A_{\mu}$. The red four-point vertex is $\delta^2 \Sigma_{\pi}/\delta A_{\mu} \delta A_{\nu}$. The double purple line is $G_{\chi}[0]$. The purple circle is $e\gamma^{\mu}/\hbar $. The purple three-point vertex is $\delta \Sigma_{\chi}/\delta A_{\mu}$, and finally the purple four-point vertex is $\delta^2 \Sigma_{\chi}/\delta A_{\mu} \delta A_{\nu}$.}
\label{fig:pfbubbles}
\end{figure}

Next, every term in Eq.~(\ref{eq:lnZ1}) is expanded up to second order in $A_{\mu}$ to obtain the total conductivity. This is described in more detail in appendix \ref{app:lnZA} and ultimately results in Eq.~(\ref{eq:dlnZ2pf}), a long expression for $\delta^2 \ln Z[A]/\delta A^{\mu} \delta A^{\nu}$. This expression contains all contributions to the total current-current correlation function, and is shown diagrammatically in Fig.~\ref{fig:pfbubbles}.
We classify the various contributions to the conductivity by the types of currents appearing inside the current-current correlation function. In the probe-fermion model, there are three contributions to the total conductivity. To see this, we recall that both the fermions in the conformal field theory, and the external probe fermion $\chi$ are coupled to a U(1) gauge field, in Eqs.~(\ref{eq:Spsi}) and (\ref{eq:Schi}), respectively. So the total charge current is the sum of these two contributions,
\begin{equation*}
J^{\mu}= J^{\mu}_{\text{cft}} + J^{\mu}_{\chi},
\end{equation*}
where $J_{\chi}^{\mu}=-ec\bar{\chi}\gamma^{\mu}\chi$ as can be seen from the free probe-fermion action in Eq.~(\ref{eq:Schi}).
The total current-current correlation function is in this case given by\footnote{The notation in the last term means we take the symmetric part of these cross terms, $A^{(\mu} B^{\nu)} = \frac{1}{2}\left(A^{\mu} B^{\nu} + A^{\nu} B^{\mu}\right)$.}
\begin{equation}\label{eq:curcurpf}\begin{aligned}
\Big\langle J^{\mu} J^{\nu} \Big\rangle &= \Big\langle  \left(J^{\mu}_{\text{cft}} + J^{\mu}_{\chi}\right)\left(J^{\nu}_{\text{cft}} + J^{\nu}_{\chi}\right)\Big\rangle \\
&= \Big\langle J^{\mu}_{\text{cft}} J^{\nu}_{\text{cft}} \Big\rangle + \Big\langle J^{\mu}_{\chi} J^{\nu}_{\chi} \Big\rangle + 2 \Big\langle J^{(\mu}_{\chi} J^{\nu)}_{\text{cft}}\Big\rangle.
\end{aligned}\end{equation}
The expectation values in Eq.~(\ref{eq:curcurpf}) are taken with respect to the full theory in Eq.~(\ref{eq:lnZ1}), i.e., including $\chi$, but with $A_{\mu}$ set to zero. Because $\chi$ is dynamical, it gives a correction proportional to $\mathcal{G}_{\chi}$ to the propagator of $\pi$ in the conformal field theory. Therefore, the conformal field theory contribution $\big\langle  J^{\mu}_{\text{cft}}J^{\nu}_{\text{cft}}\big\rangle$ can be further split up into a contribution where $\chi$ is set to zero, denoted with a subscript ``cft'' as $\big\langle J^{\mu}_{\text{cft}} J^{\nu}_{\text{cft}}\big\rangle_{\text{cft}}$ plus an additional piece that incorporates just the coupling of $\pi$ to $\chi$. This last piece is denoted by
\begin{equation*}
\Big\langle J^{\mu}_{\text{cft}} J^{\nu}_{\text{cft}} \Big\rangle_{\text{cft'}} - \Big\langle J^{\mu}_{\text{cft}} J^{\nu}_{\text{cft}} \Big\rangle_{\text{cft}}\;.
\end{equation*}
Here, the expectation value cft' is taken with respect to the partition function $Z_{\text{cft'}}$ given by
\begin{equation*}
Z_{\text{cft'}} = \int d[\bar{\pi}]d[\pi]\exp\left[\frac{i}{\hbar}S_{\text{cft}}[\bar{\pi},\pi]-i N \left(\frac{2g_1}{\hbar g_2}\right)^2 \big( \pi \big| \mathcal{G}_{\chi}[0]
\big| \pi \big)\right].
\end{equation*}
The second term in the exponent is a self-energy correction for $\pi$, which arises if we first integrate out the probe fermion $\chi$ and its conjugate $\bar{\chi}$. The conformal field theory action is given by
\begin{equation*}
S_{\text{cft}}[\bar{\pi},\pi] =\hbar N \big(\pi\big|G_{\pi}^{-1}[0]\big|\pi\big)  -i\hbar N \Tr \ln\left(-G^{-1}_{\psi}[0]\right) + i\hbar \frac{N}{2 }\Tr\ln\left(-G_{\varphi}^{-1}\right).
\end{equation*}

We will interpret the separate contributions from Eq.~(\ref{eq:curcurpf}) in the following, starting with the pure conformal field theory contribution in section \ref{sec:eleccft} and considering the additional effect of the probe fermion in section \ref{sec:elecaddchi}. As it turns out, the various contributions in Eq.~(\ref{eq:curcurpf}) may be distinguished physically by their different dependence on temperature in the dc limit.
Finally, in section \ref{sec:elecsemihol} we turn to semiholography, by means of which our model calculation of the current-current correlation function can be generalized to a probe fermion coupled to any conformal field theory.

\subsubsection{Conductivity of the conformal field theory}\label{sec:eleccft}

The conductivity of the conformal field theory is proportional to $\big\langle J^{\mu}_{\text{cft}} J^{\nu}_{\text{cft}} \big\rangle_{\text{cft}}\,$ due to charging the $\psi_i$ fermions.
The real scalar fields $\varphi_i$ are neutral.
This corresponds to taking into account diagrams (1)-(3) of Fig.~\ref{fig:pfbubbles}.

Ignoring the $\chi$-dependent contributions and the $A_{\mu}$-independent determinant of the scalar Green's function, Eq.~(\ref{eq:lnZ1}) becomes
\begin{equation}\label{eq:lnZ3}
\ln Z[A]= N \Tr\ln \Big(-G_{\psi}^{-1}[A]\Big) +\Tr\ln\Big(-N G_{\pi}^{-1}[A]\Big).
\end{equation}
The relative factor of $N$ in front of the terms in Eq.~(\ref{eq:lnZ3}) is important. It indicates that up to leading order in $N$, the conductivity is given by the free fermion conductivity. This is represented by diagram (1) in Fig.~\ref{fig:pfbubbles}. Interaction corrections, which correspond to diagrams (2) and (3), appear as $\mathcal{O}(1)$ corrections through $G_{\pi}$. The fermion self-energy contributes only at $\mathcal{O}(1/N)$ so it does not appear here.
This means that, to leading order, the fermionic conductivity of the probe-fermion model is the conductivity of $N$ species of free Dirac fermions, which is  at zero chemical potential and zero temperature given by \cite{Dirac14}
\begin{equation*}
\sigma_{\psi}(\omega) = \frac{ N e^2 |\omega|}{12 \pi \hbar c} + \mathcal{O}\left(1\right).
\end{equation*}
This is the leading fermionic contribution to the conductivity of the conformal field theory at zero temperature.

In the nonzero-temperature case, the electrical conductivity is given by the $T>0$ conductivity of the free $\psi_i$ fermions \cite{Dirac14}.
It is given by
\begin{equation}\label{eq:sigmapsi}
\sigma_{\psi}(\omega) = \frac{N e^2}{3\hbar c}\left[ \frac{\pi k_B T}{3 \hbar}\, \delta\left(\frac{\hbar\omega}{k_B T}\right) + \frac{\omega}{4\pi} \tanh\left(\frac{\hbar\omega}{4k_B T}\right)
\right],
\end{equation}
The factor $\tanh(\hbar\omega/4k_B T)$ comes from the difference of two Fermi distributions. This term is called the interband contribution and describes the contribution of particle-antiparticle excitations to transport.
The conductivity also has a delta-function peak with a weight proportional to $k_B T/\hbar$ at the point $\hbar\omega/k_B T=0$, which corresponds to intraband transport of thermally excited degrees of freedom.
 This delta-function peak is a consequence of the fact that in the probe-fermion model the conformal field theory is free at leading order in $N$. It comes from the imaginary part of a pole at zero frequency.

However, for a conformal field theory such as the theory in Ref.~\cite{Kovtun08} that is strongly interacting in the large-$N$ limit, we expect a finite result for the dc conductivity. In general, if the free result contains such a delta-function peak, interactions have the effect of smearing it out, forming a Drude-like peak in terms of a more general dimensionless function $f$ of $\hbar\omega/k_B T$, which goes to a constant in the limit $\hbar\omega/k_B T \rightarrow 0$. Schematically,
 \begin{equation*}
 \frac{k_B T}{\hbar}\delta\left(\frac{\hbar\omega}{k_B T}\right) \;\; \xrightarrow{\text{interactions}} \;\;
 \frac{k_B T}{\hbar} f\left(\frac{\hbar\omega}{k_B T}\right).
 \end{equation*}
 So in the strongly coupled case, we expect such an intraband term to contribute a term $N e^2 k_B T \,f(0)/3\hbar^2 c$ in the dc limit, where $f(0)$ is a finite and nonzero universal constant.
See Ref.~\cite{sachdev13} for an example of work where the universal coefficients of current-current correlation functions of a certain conformal field theory are computed in a $1/N$ expansion.

In this way, the probe-fermion model reproduces the result in Ref.~\cite{Kovtun08}, namely that the dc conductivity of a conformal field theory in a thermal state is linear in temperature, in the (3+1)-dimensional case. This is to be expected from dimensional analysis, as temperature is the only scale present. Depending on whether the theory is interacting or free, the prefactor can be finite, as in Ref.~\cite{Kovtun08}, or infinite, when the linear scaling in temperature comes from a delta-function peak.

\subsubsection{Additional effects of the probe fermion}\label{sec:elecaddchi}

When the probe fermion is added to the conformal field theory, this gives rise to the rest of the terms in Eq.~(\ref{eq:curcurpf}), namely,
\begin{equation}\label{eq:curcurpfv2}\begin{aligned}
\Big\langle J^{\mu} J^{\nu} \Big\rangle - \Big\langle J^{\mu}_{\text{cft}} J^{\nu}_{\text{cft}} \Big\rangle_{\text{cft}}&= \Big\langle J^{\mu}_{\chi} J^{\nu}_{\chi} \Big\rangle + 2 \Big\langle J^{(\mu}_{\chi} J^{\nu)}_{\text{cft}}\Big\rangle \\&+ \Big\langle J^{\mu}_{\text{cft}} J^{\nu}_{\text{cft}} \Big\rangle_{\text{cft'}} - \Big\langle J^{\mu}_{\text{cft}} J^{\nu}_{\text{cft}} \Big\rangle_{\text{cft}}\;.
\end{aligned}\end{equation}
All these terms contribute at $\mathcal{O}(1)$ to the total conductivity.

The first term on the right-hand side of Eq.~(\ref{eq:curcurpfv2}), i.e., $\big\langle J^{\mu}_{\chi} J^{\nu}_{\chi} \big\rangle$, is the current-current correlation function of the probe fermion, and it is represented by diagram (4) of Fig.~\ref{fig:pfbubbles}.
 Its contribution to the electrical conductivity is analogous to the conductivity of the interacting Dirac fermion in semiholography, computed in Ref.~\cite{Dirac14}. So as in Eq.~(\ref{eq:dscond}), the dc conductivity of the probe fermion is proportional to $T^{3-4M}$, with $-1/2<M<1/2$.
 This can be understood by dimensional analysis.
  In the strong-coupling limit of the conformal field theory, the only dimensionful quantities present are the temperature and the coupling between probe fermion and conformal field theory, called $g$ in the probe-fermion model.
 For fixed and finite $g$, every occurrence of the self-energy provides a power $2M$ of temperature, while every propagator of $\chi$ contributes the inverse power. This is because in the dc limit, the self-energy dominates over the free part of the $\chi$ propagator. So by counting propagators and comparing it to the dimension of conductivity in 3+1 dimensions, we can determine the scaling behavior of these diagrams in the dc limit.

The second term on the right-hand side of Eq.~(\ref{eq:curcurpfv2}), i.e., $\big\langle J^{(\mu}_{\chi} J^{\nu)}_{\text{cft}}\big\rangle$, is an interference term between the probe-fermion current and the current of the conformal field theory. This term gives rise to diagrams (5) and (6) of Fig.~\ref{fig:pfbubbles}. Because the vertex function contains  a factor of $\Sigma_{\chi}$, the above counting suggests that this term scales with temperature as $T^{2-2M}$.

 Thirdly, the combination $\big\langle J^{\mu}_{\text{cft}} J^{\nu}_{\text{cft}} \big\rangle_{\text{cft'}} - \big\langle J^{\mu}_{\text{cft}} J^{\nu}_{\text{cft}} \big\rangle_{\text{cft}}$ represents the effect that the addition of the probe fermion has on the current-current correlation function of the conformal field theory. This contribution is represented by diagrams (7) and (8). In this case, dimensional analysis suggests that these terms scale again linearly with temperature.

 Summarizing, based on the discussions above and in section \ref{sec:eleccft}, we conclude that including all contributions in Fig.~\ref{fig:pfbubbles}, i.e., up to subleading order in $N$, the total electrical conductivity in the dc limit is of the form
 \begin{equation*}
 \sigma_{dc} \sim \underbrace{\Big(\mathcal{O}\left(N\right)+ \mathcal{O}\left(1\right)\Big)T}_{\text{conformal field theory}} + \underbrace{\mathcal{O}\left(1\right) T^{2-2M}}_{\text{interference}} +
\underbrace{\mathcal{O}\left(1\right)T^{3-4M}}_{\text{probe fermion}} + \mathcal{O}\left(\frac{1}{N}\right).
 \end{equation*}
It is interesting to note that in the probe-fermion model, the probe-fermion conductivity and interference contribution can be identified from their anomalous scaling with temperature, despite the fact that they are subleading in $N$.

We note one more thing on this classification of diagrams.
The total action in Eq.~(\ref{eq:Stotpfmodel}) has a local U(1) symmetry associated with conservation of the total charge current $J^{\m}$. The procedure to compute the current-current correlation function from appendix \ref{app:lnZA} leading to the diagrams in Fig.~\ref{fig:pfbubbles} does not violate this gauge invariance. As a consequence, the Ward identity for current conservation is satisfied and $\big\langle J^{\mu} J^{\nu}\big\rangle$ is automatically transversal, i.e., \footnote{To be precise, the finite part of $\big\langle J^{\mu} J^{\nu}\big\rangle$ is transversal. There may be longitudinal divergent parts that drop out after regularization.}
\begin{equation*}
\dau_{\m}\Big\langle J^{\mu} J^{\nu}\Big\rangle=0.
\end{equation*}
It is interesting to consider how precisely the various contributions cancel when taking the divergence of $\big\langle J^{\mu} J^{\nu}\big\rangle$. For instance, the divergence of diagram (1) vanishes by itself because it contains only bare propagators and bare vertices. However, the probe-fermion contribution, diagram (4), contains dressed propagators but bare vertices, and therefore we need additional contributions to satisfy the Ward identity. These additional contributions are supplied by diagrams (5), (6) and (7).
The divergence of diagrams (2), (3) and (8) vanishes because of the conformal field theory Ward identity in the absence of $\chi$.

 In Ref.~\cite{Dirac14}, diagram (4) was computed numerically and its physical properties discussed in details.
  According to the classification in section \ref{sec:elecft}, diagram (4) is the only contribution to the fermionic conductivity. The conductivity contributions due to the dependence of the self-energy on the gauge field can be interpreted as interference with the conformal field theory, and a correction to the conductivity of the conformal field theory.

\subsubsection{Conductivity in semiholography}\label{sec:elecsemihol}

It is enlightening to reconsider the electrical conductivity of the boundary theory in the context of semiholography. The boundary action of the dual
conformal field theory coupled to a fermionic dynamical source $\chi$ resembles the action in Eq.~(\ref{eq:critfermions}), namely, it is given by
\begin{equation}\label{eq:semiholoaction}
S = S_{\text{cft}}[\bar{O},O] + S_0[\bar{\chi},\chi]+i g \int d^4 x \,\left(\bar{\chi} O + \bar{O} \chi\right).
\end{equation}
Here, $O$ is a fermionic composite operator in the conformal field theory $S_{\text{cft}}$. In general, in the presence of a boundary gauge field $A_{\m}$ which couples to both the conformal field theory and to the dynamical source $\chi$, this is modified to
\begin{equation*}
S = S_{\text{cft}}[\bar{O},O;A]+\big(\chi\big|\hbar G^{-1}_0-e\slashed{A}\big|\chi\big)+ig \big(\chi\big|O\big)+ig\big(O\big|\chi\big),
\end{equation*}
where
\begin{equation*}
S_{\text{cft}}[\bar{O},O;A] = S_{\text{cft}}[\bar{O},O] + \frac{1}{c}\int d^4 x \, J^{\mu}_{\text{cft}} \,A_{\mu},
\end{equation*}
and $J^{\mu}_{\text{cft}}$ is again the current of the conformal field theory. The conformal field theory can be integrated out, and for the dynamical source $\chi$, this gives rise to an effective self-energy,
\begin{equation*}
\Sigma_{\chi}[A](x,x')=-i \left(\frac{g}{\hbar}\right)^2 \Big\langle \bar{O}(x') O(x)\Big\rangle_{\text{cft}+A},
\end{equation*}
and the subscript cft+$A$ indicates that this is the expectation value with respect to the conformal field theory action in the presence of a current $J^{\mu}_{\text{cft}}$ sourced by the boundary gauge field.

\begin{figure}[h!]
\vskip 10pt
\centering
\includegraphics[width=0.75\textwidth]{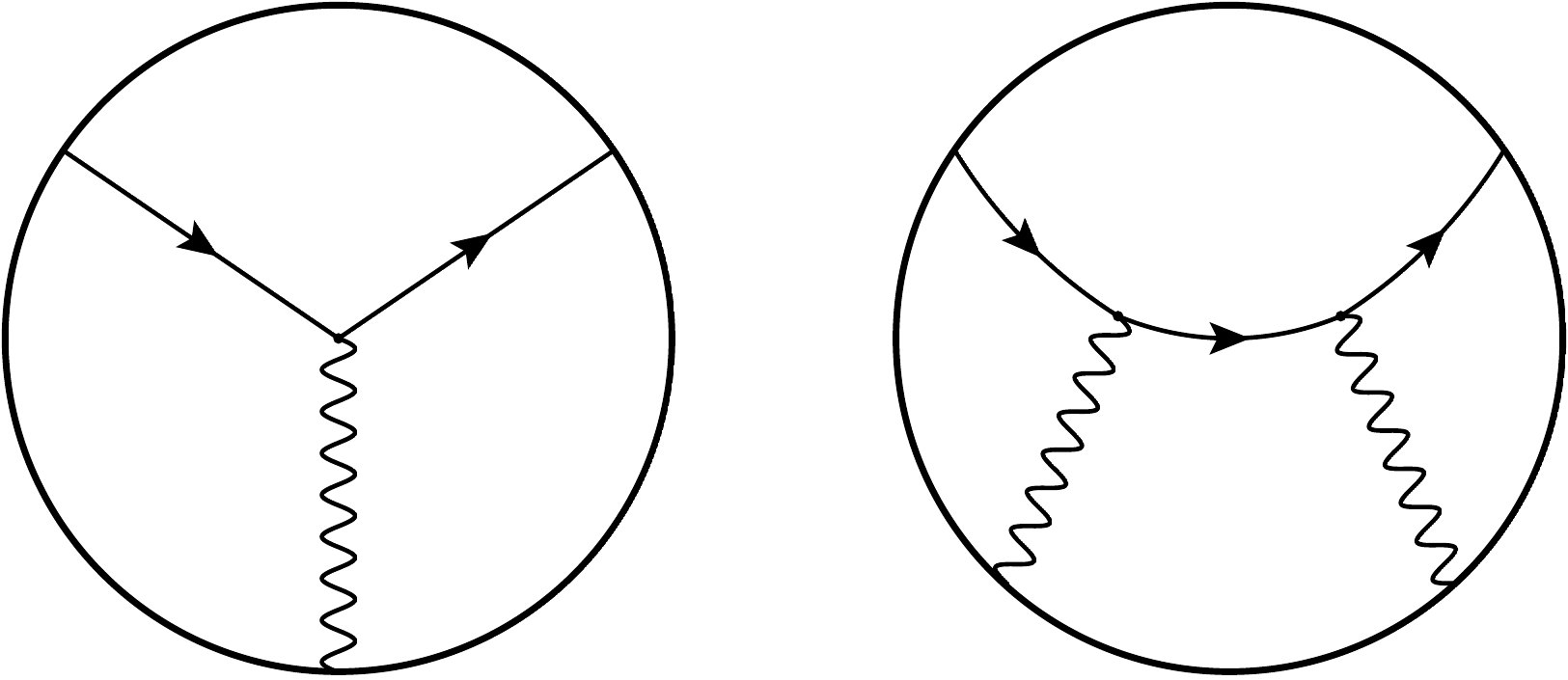}
\caption{Witten diagrams for the correlators $\big\langle \bar{O} O J^{\mu}_{\text{cft}}\big\rangle_{\text{cft}}$ and $\big\langle \bar{O} O J^{\mu}_{\text{cft}}J^{\nu}_{\text{cft}}\big\rangle_{\text{cft}}$.}
\label{fig:witten}
\end{figure}

 As before, the total electrical conductivity of the boundary theory is proportional to the total current-current correlation function, which is in turn  proportional to the expression $\delta^2 \ln Z[A]/\delta A_{\mu}\delta A_{\nu}$. It is convenient to integrate out the $\bar{\chi}$ and $\chi$ fields as well and, for taking these functional derivatives, to expand the generating functional $\ln Z[A]$ up to second order in the gauge field. The same procedure was carried out for the probe-fermion model in appendix \ref{app:lnZA}. Besides the pure conformal field theory contribution, the resulting expression contains the various purple diagrams in Fig.~\ref{fig:pfbubbles} built from the $\chi$ propagators and vertices present in the theory. In particular, there will be vertices of the form $\delta \Sigma_{\chi}/\delta A_{\mu}$ and $\delta^2 \Sigma_{\chi}/\delta A_{\mu}\delta A_{\nu}$. To investigate these, the self-energy of the dynamical source is expanded in powers of the gauge field. We have
\begin{equation*}\begin{aligned}
\Big\langle \bar{O}(x) O(x')\Big\rangle_{\text{cft}+A} & \simeq\Big\langle \bar{O}(x) O(x')\Big\rangle_{\text{cft}} + \frac{i}{\hbar c}\int d^4 y \Big\langle \bar{O}(x)O(x') J^{\mu}_{\text{cft}}(y)\Big\rangle_{\text{cft}}\, A_{\mu}(y) \\
&+ \left(\frac{i}{\hbar c}\right)^2 \int d^4 y \int d^4y' \Big\langle \bar{O}(x)O (x') J^{\mu}_{\text{cft}}(y) J^{\nu}_{\text{cft}}(y')\Big\rangle_{\text{cft}} \, A_{\mu}(y) A_{\nu}(y').
\end{aligned}\end{equation*}
This leads to the following expression for the three-point and four-point vertices
\begin{equation}\label{eq:vertexholo}\begin{aligned}
& \frac{\delta \Sigma_{\chi}}{\delta A_{\mu}} = \frac{1}{\hbar c} \left(\frac{g}{\hbar}\right)^2\Big\langle \bar{O} O J^{\mu}_{\text{cft}}\Big\rangle_{\text{cft}} \,,\\
&\frac{\delta^2\Sigma_{\chi}}{\delta A_{\mu}\delta A_{\nu}} = i\left(\frac{g}{\hbar^2 c}\right)^2 \Big\langle \bar{O}O J^{\mu}_{\text{cft}} J^{\nu}_{\text{cft}}\Big\rangle_{\text{cft}} \,,
\end{aligned}\end{equation}
respectively. Up to here this was just a field-theory calculation, but now we connect this result to holography. The vertices on the left-hand side of Eq.~(\ref{eq:vertexholo}) contribute i.a. to the interference terms needed to satisfy the Ward identity.\footnote{We have shown this in the probe-fermion model in section \ref{sec:elecaddchi}, but it is not hard to see that this is a general result.} On the right-hand side we find precisely the correlators of operators in a conformal field theory with a holographic dual. So in a semiholographic theory, the full conductivity satisfying the Ward identity can be computed with holography. This amounts to computing, besides the self-energy that is proportional to $\big\langle \bar{O} O\big\rangle_{\text{cft}}$, the correlators $\big\langle \bar{O} O J^{\mu}_{\text{cft}}\big\rangle_{\text{cft}}$ and $\big\langle \bar{O} O J^{\mu}_{\text{cft}}J^{\nu}_{\text{cft}}\big\rangle_{\text{cft}}$ using the holographic prescription \cite{umutpanos}. These correlators can be represented by the Witten diagrams in Fig~\ref{fig:witten}.

As a final remark, we note that the coupling in the action in Eq.~(\ref{eq:critfermions}) is a special case of the coupling in the semiholographic action in Eq.~(\ref{eq:semiholoaction}), namely, in Eq.~(\ref{eq:critfermions}), the operator $O$ is given by the product of the elementary fermion and a critical boson, $O = \chi \Phi$. Because the coupling in Eq.~(\ref{eq:semiholoaction}) contains only one power of $\chi$, integrating out the critical theory leads to an effective self-energy for $\chi$, i.e., a term in the action proportional to $g^2 \bar{\chi}\chi\langle \bar{O}O\rangle_{\text{cft}}$. In contrast, integrating out the critical boson $\Phi$ in Eq.~(\ref{eq:critfermions}) yields a nonlocal effective interaction for $\chi$ of the form $g^2\left(\bar{\chi}\chi \right)^2 \langle \Phi^* \Phi\rangle_{\text{cft}}.$\footnote{Such an effective four-point interaction is the starting point of the Fock model introduced in the next section.} This effective interaction can lead, for instance in the Hartree-Fock approximation, to an effective self-energy for $\chi$. In a semi-holographic setting, we need a bulk scalar field dual to the order parameter $\Phi$ that is sourced by the fermion bilinear $\bar{\chi}\chi$ on the boundary. This results in the required interaction term for $\chi$ where $\langle \Phi^* \Phi\rangle_{\text{cft}}$ is related to the bulk-to-boundary propagator of the dual scalar field. As Eq.~(\ref{eq:critfermions}) just serves as a motivation for the problem discussed in this paper, a detailed treatment of this case in semiholography is beyond the scope of this paper.

\section{Fock model}\label{sec:fock}
In section \ref{sec:pfmodel} we have seen that the probe-fermion model offers an interpretation of the dynamical-source model. Here the dynamical source is analogous to a probe fermion coupled to a conformal field theory.
A disadvantage of the probe-fermion model is the fact that for $\eta=0$, we do not recover the $M=1/2$ behavior with logarithmic corrections seen in the dynamical-source model. Furthermore, the conformal field theory of the probe-fermion model is free in the large-$N$ limit. As a consequence the dominant contribution to the electrical conductivity comes from free $\psi_i$ fermions, and it thus diverges in the dc limit.

 These problems are addressed in the second field-theory model that we consider, the Fock model. Here, there is no separate probe fermion. Instead, we consider fermions with an effective interaction due to a propagator containing an anomalous dimension. This effective interaction reduces to a Coulomb-like potential for $\eta=0$, which is the desired behavior in the $M=1/2$ case of the dynamical-source model. A nontrivial self-energy, given by the selfconsistent Fock diagram, comes about by introducing a bosonic and nonlocal auxiliary field. In the strong-coupling limit, the fermionic propagator reduces to the inverse of the self-energy, and this suggests that the kinetic term $G_0^{-1}$ in the propagator arises due to finite-coupling corrections. Thus, we have a second interpretation of the dynamical-source model, where the source fermion lives in the dual conformal field theory, and the dynamical part of its propagator is an effective way to incorporate finite-coupling corrections.

In the Fock model, we have again $N$ species of Dirac fermions.
They are minimally coupled to a gauge field $A_{\m}$ and
interact via a nonlocal, real interaction potential $\Delta$. The partition function is
\begin{equation}\label{eq:ZAfock}
Z[A]=\int d[\bar{\psi}]d[\psi] \exp\bigg[\frac{i}{\hbar} \Big(S_0[\bar{\psi},\psi;A] + S_{\Delta}[\bar{\psi},\psi]\Big)\bigg],
\end{equation}
where the actions are given by
\begin{equation}\label{eq:S0fock}
S_0[\bar{\psi},\psi;A]=-i \hbar \sum_{i=1}^N \int d^4 x \,\bar{\psi}_i\Big(\slashed{\dau}-\frac{ie}{\hbar}\slashed{A}\Big)\psi_i,
\end{equation}
and
\begin{equation*}\begin{aligned}
S_{\Delta}[\bar{\psi},\psi] =\frac{\hbar g}{2N}
\sum_{i,i'=1}^N \int d^4 x \int d^4 x' \Delta(x-x')\,\bar{\psi}_i(x)\,\psi_{i'}(x) \,\bar{\psi}_{i'}(x')\, \psi_i(x').
\end{aligned}\end{equation*}
As discussed before, the motivation for this form of $S_{\Delta}$ is that a nonlocal interaction for the fermions is required to make their self-energy nontrivial in the large-$N$ limit. For a microscopic origin of this form of the interaction $\Delta(x-x')$, we imagine that integrating out other fields in a more complicated theory could lead to such a nonlocal effective interaction term, as was already discussed at the end of section \ref{sec:elecsemihol}. Because of the nonlocality, this interaction can carry momentum, and we choose it to have the momentum-dependence of a relativistic scalar with an anomalous dimension $\eta$,
\begin{equation}\label{eq:Delta}
\Delta(x-x')=\int\frac{d^4 k}{(2\pi)^4} \frac{1}{-k^{2-\eta}} e^{i k\cdot (x-x')}.
\end{equation}
We proceed by performing a Hubbard-Stratonovich transformation that decouples the interaction, by multiplying the partition function by
\begin{equation}\label{eq:HSrho}\begin{aligned}
1 &= \int d[\rho]\exp\bigg[\frac{i}{\hbar}\int d^4 x \int d^4 x' \sum_{\alpha,\beta}\Big(\rho_{\beta\alpha}(x',x) - \sum_{i=1}^N \frac{\hbar g }{N} \bar{\psi}_{i,\alpha}(x)\psi_{i,\beta}(x')\Big)\\
&\times \frac{N \Delta(x-x')}{2\hbar g } \Big(\rho_{\alpha\beta}(x,x')-\sum_{i'=1}^N\frac{\hbar g}{N} \bar{\psi}_{i',\beta}(x') \psi_{i',\alpha}(x)\Big)\bigg].
\end{aligned}\end{equation}
We have written down the spin indices $\alpha,\beta$ explicitly. The auxiliary field $\rho$ is a matrix in these indices, is nonlocal, and its expectation value is proportional to the fermionic density matrix. Defining the collective field $\rho$ in this way is indicative of Fock theory, which explains the name of this model. As in the probe-fermion model, the factors of $N$ are chosen such that certain diagrams survive in the large-$N$ limit.
We introduce  new shorthand notation, by means of which the expression in Eq.~(\ref{eq:HSrho}) is written as
\begin{equation*}
1= \int d[\rho]\exp\bigg[\frac{i}{\hbar}\sum_{i,i'=1}^N\big(\rho+\frac{\hbar g}{N} \psi_i\bar{\psi}_i \big|\big| \frac{N \Delta}{2\hbar g} \big|\big| \rho+\frac{\hbar g}{N} \psi_{i'}\bar{\psi}_{i'} \big)\bigg].
\end{equation*}
Compared to the notation in section \ref{sec:pfmodel}, the double lines now express the dependence on two spacetime points of the field $\rho$ and the product $\bar{\psi}\psi$. After this transformation, the action for the fermions is quadratic, and we can read off the inverse Green's function for a single fermion species. It is a functional of $\rho$ and $A_{\mu}$, and its matrix elements are in the coordinate representation given by
\begin{equation*}\begin{aligned}
&G^{-1}_{\psi,\alpha\beta}[A,\rho](x, x')=-i \left(\slashed{\dau}_{\alpha\beta}-\frac{ie}{\hbar}\slashed{A}_{\alpha\beta}(x)\right)\delta^4 (x-x')  -\frac{1}{\hbar}\rho_{\alpha\beta}(x,x')\Delta(x-x').
\end{aligned}\end{equation*}
From now on, the spinor indices are again suppressed in our notation. We can integrate out the fermions, which results in the partition function
\begin{equation}\label{eq:Zrho}
Z[A]= \int d[\rho] \exp\bigg[N \Tr \ln\Big(-G_{\psi}^{-1}[A,\rho]\Big)+N\big(\rho\big|\big| \frac{i}{\hbar} \frac{\Delta}{2\hbar g}\big|\big|\rho\big)\bigg].
\end{equation}
We do a fluctuation expansion of $\rho$ around its expectation value by writing $\rho=\langle\rho\rangle+\delta \rho$. In the large-$N$ limit, the approximation where we take only the leading term into account, becomes exact, because the path integral is dominated by this value. In this case, it is the mean-field approximation where we replace $\rho$ by $\langle \rho \rangle$. The field $\rho$ is bosonic so its expectation value can be nonzero. Setting the first-order terms in $\delta \rho$ to zero leads to an equation for $\langle\rho\rangle$. First, we expand the fermionic propagator to first order in $\delta \rho$,
\begin{equation*}\begin{aligned}
G^{-1}_{\psi}[A,\rho](x,x')&= G_{\psi}^{-1}[A,\langle \rho\rangle](x,x')- \frac{1}{\hbar} \Delta(x-x') \delta\rho(x,x'),
\end{aligned}\end{equation*}
where we have defined the inverse fermion propagator in the mean-field approximation as
\begin{equation}\label{eq:Gpsimf}\begin{aligned}
G_{\psi}^{-1}[A,\langle \rho\rangle](x, x') &= -i \left(\slashed{\dau}-\frac{ie}{\hbar}\slashed{A}(x)\right)\delta^4(x-x') - \frac{1}{\hbar}\big\langle\rho(x,x')\big\rangle \Delta(x-x').
\end{aligned}\end{equation}
This is the Dyson equation for the dressed fermion propagator. The second term on the right-hand side of Eq.~(\ref{eq:Gpsimf}) is the self-energy of the fermions, which is proportional to $\langle \rho \rangle \Delta$. Thus, the mean-field approximation to $\rho$ leads to the Fock approximation to the fermion self-energy, which becomes exact for large $N$. This is in contrast to the probe-fermion model, which can be seen as a Hartree-like theory, in which the auxiliary field is local and the self-energy of the $\psi_i$ fields is $1/N$ suppressed.

\begin{figure}[h!]
\vskip 10pt
\centering
\includegraphics[width=0.95\textwidth]{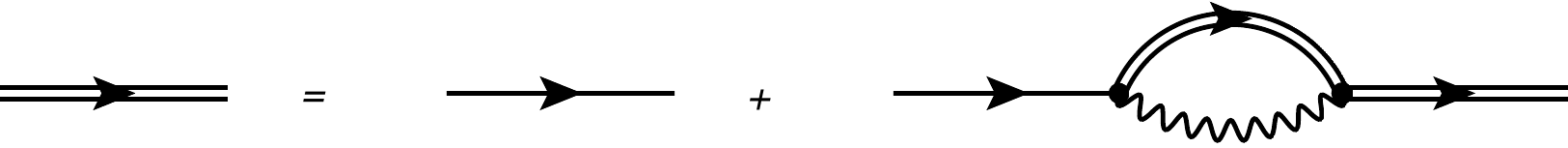}
\caption{Graphical representation of the Dyson equation from Eq.~(\ref{eq:dysonpsi}). Double lines denote the full Green's function $G_{\psi}$, while single lines denote the free propagator $1/\slashed{k}$. The wiggly line is the interaction $\Delta$. The Fock bubble comes with a factor $g$, or equivalently, two factors $\sqrt{g}$. The latter can be associated with the solid dots.}
\label{fig:fock_dyson}
\end{figure}

To find the fermion self-energy in the Fock model, we first set $A_{\mu}=0$ so that $\langle\rho\rangle$ depends only on the difference in coordinates. Then we expand the logarithm in Eq.~(\ref{eq:Zrho}) in fluctuations, and demanding the first-order terms in $\delta \rho$ in the partition function to vanish, leads to the following equation for $\langle \rho \rangle$,
\begin{equation}\label{eq:rhoexp}
\big\langle \rho(x-x')\big\rangle= \frac{g\hbar}{i} G_{\psi}[0,\langle \rho\rangle](x-x').
\end{equation}
Together, Eq.~(\ref{eq:Gpsimf}) and Eq.~(\ref{eq:rhoexp}) form a recursive equation for the dressed fermion propagator, and thus for the fermionic self-energy. This recursive equation is elegantly expressed in momentum space as
\begin{equation}\label{eq:dysonpsi}
G^{-1}_{\psi}(k)= \slashed{k} -\Sigma_{\psi}(k),
\end{equation}
with the self-energy
\begin{equation*}
\Sigma_{\psi}(k)=- ig\int \frac{d^4 q}{(2\pi)^4} \Delta(k-q)G_{\psi}(q).
\end{equation*}
Note that we have also suppressed the functional dependence on $A$, $\rho$ and $\langle \rho \rangle$, so that the dressed Green's function is denoted just by $G_{\psi}$. This Dyson equation for the fermion propagator is shown diagrammatically in Fig.~\ref{fig:fock_dyson}.

\subsection{Fermionic Fock self-energy}\label{sec:fockself}
Finding the fermionic self-energy requires solving the recursive equation Eq.~(\ref{eq:dysonpsi}), which cannot be solved analytically. However, we can solve it both in the free case and the very strong-coupling limit. In the first, we just set $g=0$ so that the self-energy piece drops out, and obtain the trivial result $G_{\psi}(k)=1/\slashed{k}$. We can depart from this trivial value of the coupling by means of perturbation theory in $g$.

The strong-coupling limit is the opposite limit, with the effect that the self-energy term dominates over the $\slashed{k}$ term in Eq.~(\ref{eq:dysonpsi}), so that we can ignore it, and Eq.~(\ref{eq:dysonpsi}) becomes
\begin{equation}\label{eq:dysonstrong}
\frac{1}{i g}G_{\psi}^{-1}(k) = \int \frac{d^4 q}{(2\pi)^4} \Delta(k-q) G_{\psi}(q).
\end{equation}
In this limit, the fermion propagator must depend on $g$, otherwise Eq.~(\ref{eq:dysonstrong}) is inconsistent. It has dimensions of length in our conventions. Furthermore, in the absence of scales such as temperature, it must be a single power of $k$, it is fermionic so there is also a $\slashed{k}$ involved, and of course there can be a numerical prefactor. So we plug in the following Ansatz,
\begin{equation*}
G_{\psi}(k)= g^y k^{x-1} \slashed{k} h(\eta),
\end{equation*}
where the function $h(\eta)$ is dimensionless.
Plugging this Ansatz into Eq.~(\ref{eq:dysonstrong}) and equating the powers of $g$ on both sides of the equality, leads to $y=-1/2$, and consequently, $x=-1-\eta/2$. Then, the Green's function in the strong-coupling limit is given by
\begin{equation}\label{eq:fockGpsi}
G_{\psi}(k)=-\frac{1}{\Sigma_{\psi}(k)}=\frac{\slashed{k} \,h(\eta)  }{ \sqrt{g} k^{2+\frac{\eta}{2}}}.
\end{equation}
We can solve $h(\eta)$ from Eq.~(\ref{eq:dysonstrong}) by doing the momentum integral on the right-hand side, which is described in more detail in appendix \ref{app:fockloop}, and results in
\begin{equation}\label{eq:hfock}
\frac{1}{h^2(\eta)}= \frac{\Gamma(1-\frac{\eta}{4})}{(4\pi)^2\Gamma(1+\frac{\eta}{4})} \left[\frac{\Gamma(1+\frac{\eta}{2})\Gamma(-\frac{\eta}{4})}{\Gamma(1-\frac{\eta}{2})\Gamma(2+\frac{\eta}{4})}-\frac{4+\eta}{2\eta}\,\frac{\Gamma(-1-\frac{\eta}{4})\Gamma(2+\frac{\eta}{2})}{\Gamma(3+\frac{\eta}{4})\Gamma(-\frac{\eta}{2})}\right].
\end{equation}
In the strong-coupling limit the Green's function is minus the inverse of the self-energy. The self-energy of the fermions is thus given by
\begin{equation}\label{eq:selfv3}
\Sigma_{\psi}(k) = -\frac{\sqrt{g} }{h(\eta)} \slashed{k} k^{\frac{\eta}{2}}.
\end{equation}
Note that this result is proportional to $\sqrt{g}$, which expresses the
nonperturbative nature of the calculation. Switching to the retarded prescription amounts to setting $ck^0 = \omega+i0$. Thus, in the strong-coupling limit $G_{\psi}$ coincides with the result from Eq.~(\ref{eq:selfv0}) in the dynamical-source model, if we make the following identifications,
\begin{equation*}\begin{aligned}
&\eta=4M-2,\\
&\sqrt{g}=\frac{\lambda \,\Gamma(\frac{1}{2}-M) }{c\,4^M \Gamma(\frac{1}{2}+M)  }h\left(4M-2\right).
\end{aligned}\end{equation*}
We see that the range $-2<\eta<0$, where $h(\eta)$ is real, covers only the positive-$M$ part of the self-energy in the dynamical-source model, as it corresponds to $0<M<1/2$.
However, in this case the value $\eta=0$ does correspond to $M=1/2$.

\subsection{Electrical conductivity}\label{sec:elecfock}
To compute the electrical conductivity of the fermions in the Fock model, we can start from the generating functional $\ln Z[A]$, which is, after making use of Eq.~(\ref{eq:rhoexp}) in the presence of $A_{\mu}$, given by
\begin{equation*}
\ln Z[A] =N \Tr\ln\Big(-G^{-1}_{\psi}[A]\Big) -i  N  \, \,\big(G_{\psi}[A]\,\big|\big|\frac{g\Delta}{2}\big|\big|\,G_{\psi}[A]\,\big).
\end{equation*}
In principle, the procedure is the same as for the probe-fermion model, i.e., taking two functional derivatives of $\ln Z[A]$ with respect to the gauge field.
However, less work is required if we make use of the fact that, in the Fock model, the total current density operator is exactly known in terms of the fermionic fields. It is given by
\begin{equation*}
J^{\mu}(x) = -ec\sum_{j=1}^N \psib_j(x)  \gamma^{\mu}\psi_j(x).
\end{equation*}
The current-current correlation function is therefore proportional to
\begin{equation}\label{eq:Pimnv1}
\frac{\delta\big\langle J^{\mu}(x)\big\rangle}{\delta A_{\nu}(y)}\bigg|_{A=0}= Ni  c e\lim_{x'\rightarrow x} \text{tr}\left[\gamma^{\mu} \frac{\delta G_{\psi}[A](x,x')}{\delta A_{\nu}(y)}\bigg|_{A=0}\right],
\end{equation}
where the expectation value is taken with respect to the interacting action in the presence of $A_{\mu}$, and the lowercase trace is  just over spinor indices. Evaluation of Eq.~(\ref{eq:Pimnv1}) requires knowing $G_{\psi}$ up to first order in $A_{\mu}$. This information is obtained from the mean-field Dyson equation for $G_{\psi}$ which we found in section \ref{sec:fockself}, and which we recall here,
\begin{equation}\begin{aligned}\label{eq:dysoncond}
G_{\psi}^{-1}[A](x,x')
& = G_0^{-1}(x,x')  - \frac{e}{\hbar}\delta^4(x-x')\slashed{A}(x)-\Sigma_{\psi}[A](x,x'),
\end{aligned}\end{equation}
with the inverse noninteracting fermion Green's function $G_0^{-1}(x,x') = - i \slashed{\dau}\delta^4(x-x') $ and with the self-energy
\begin{equation*}
\Sigma_{\psi}[A](x,x')= \frac{g}{i} G_{\psi}[A](x,x')\Delta(x-x'),
\end{equation*}
which depends on $A_{\mu}$ through $G_{\psi}$. It is not necessary to invert this Dyson equation. Namely, using the definition of the matrix inverse in the coordinate representation, it is easy to show that
\begin{equation*}\begin{aligned}
\frac{\delta G_{\psi}[A](x,x')}{\delta A_{\nu}(y)}\bigg|_{A=0}
&= -\int d^4 y' \int d^4 y''\, G_{\psi}[0](x,y')\, \Gamma^{\nu}(y',y'';y) \, G_{\psi}[0](y'',x').
\end{aligned}\end{equation*}
We have introduced the three-point function $\Gamma^{\nu}$, which is just the functional derivative with respect to $A_{\nu}$ of the inverse Green's function, i.e.,
\begin{equation*}
\Gamma^{\nu}(y',y'';y) = \frac{\delta G_{\psi}^{-1}[A](y',y'')}{\delta A_{\nu}(y)}\bigg|_{A=0}.
\end{equation*}

\begin{figure}
\vskip 10pt
\centering
\includegraphics[width=.3\textwidth]{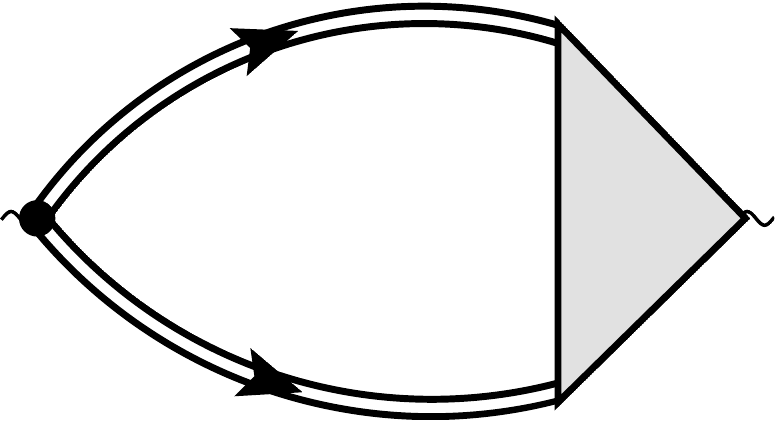}
\caption{Diagram of the current-current correlation function from Eq.~(\ref{eq:curcurfock}) containing the three-point vertex. The filled triangle is $\Gamma^{\mu}$, the solid circle denotes the bare vertex proportional to $e\gamma^{\mu}$ and the double lines represent the dressed fermion propagator $G_{\psi}[0]$. }
\label{fig:bubble3vertex}
\end{figure}

\begin{figure}
\vskip 10pt
\centering
\includegraphics[width=.7\textwidth]{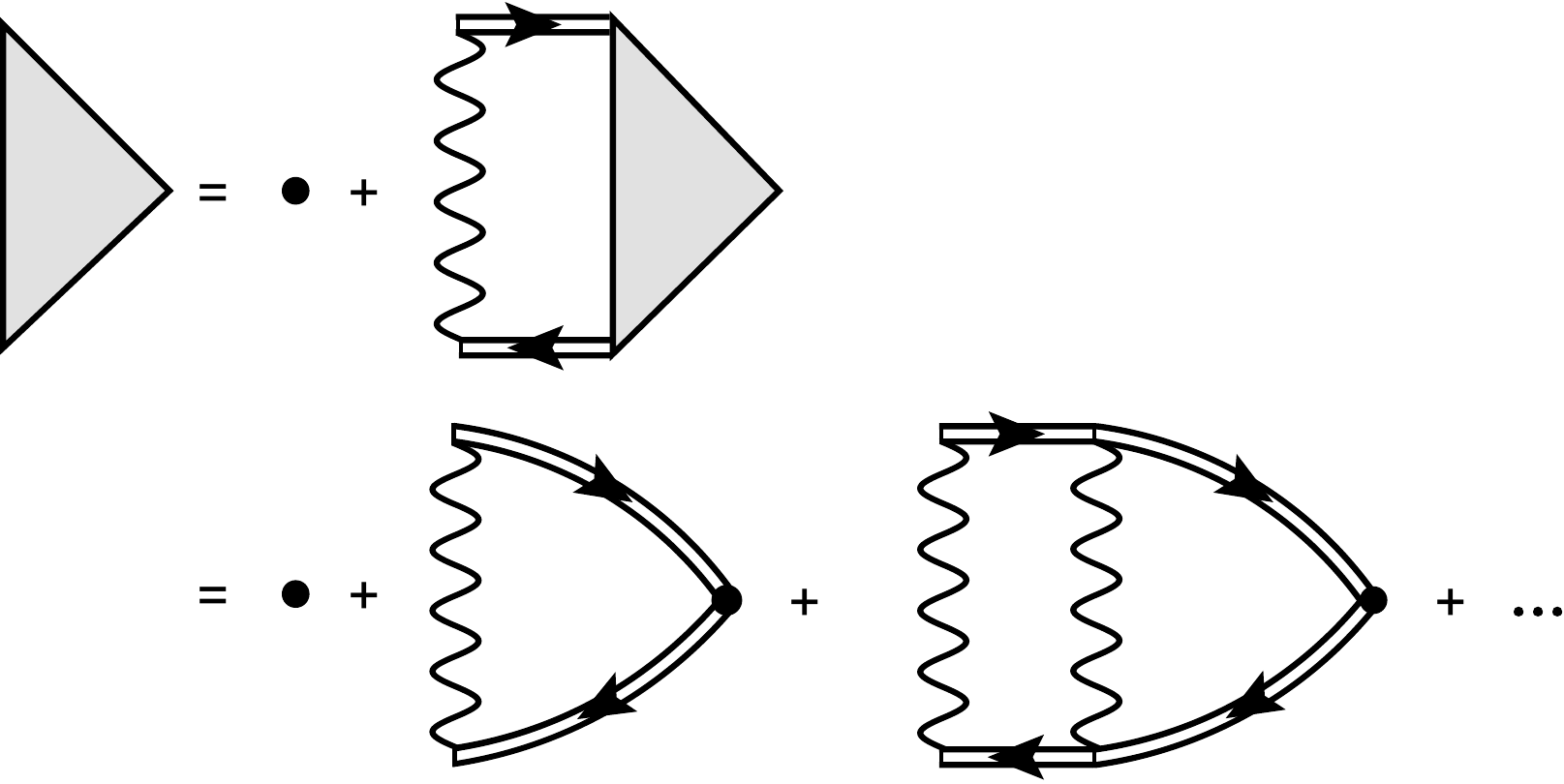}
\caption{The recursive relation for the three-point vertex $\Gamma^{\mu}$ from Eq.~(\ref{eq:Gammarec}), where we also made use of Eq.~(\ref{eq:selfderiv}). The notation is the same as in Fig.~\ref{fig:bubble3vertex}, and each wiggly line is an interaction $g\Delta$. }
\label{fig:3vertex_recu}
\end{figure}

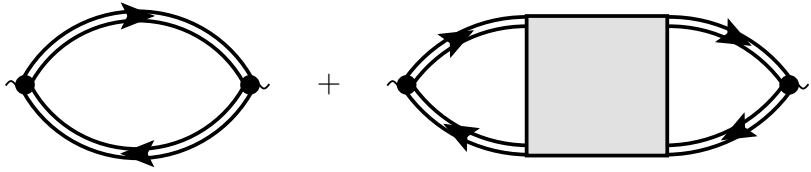
\begin{figure}
\vskip 10pt
\centering
\includegraphics[width=.7\textwidth]{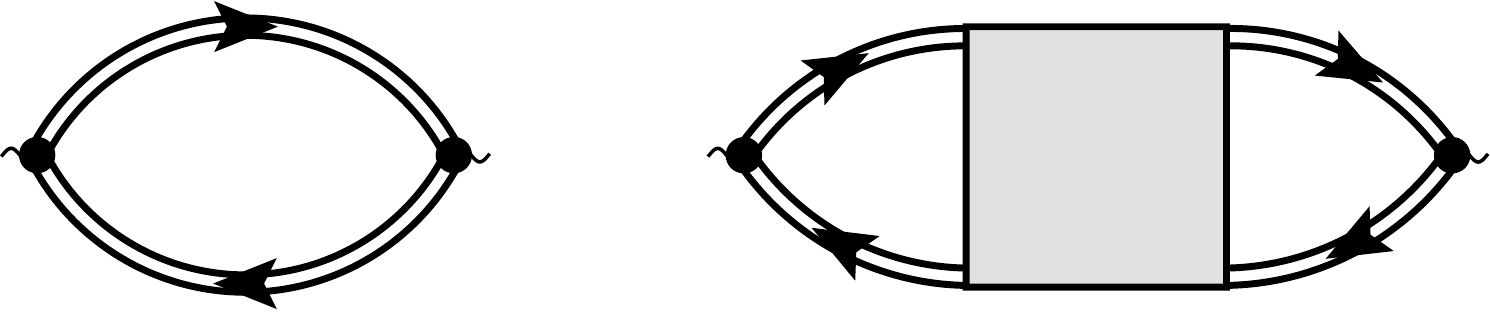}
\begin{picture}(0,0)
\put(-188,29){\large{$+$}}
\end{picture}
\caption{Diagrams for the current-current correlation function in terms of the four-point vertex in Fig.~\ref{fig:4vertex_recu}. }
\label{fig:bubble4vertex}
\end{figure}

\begin{figure}
\vskip 10pt
\centering
\includegraphics[width=.6\textwidth]{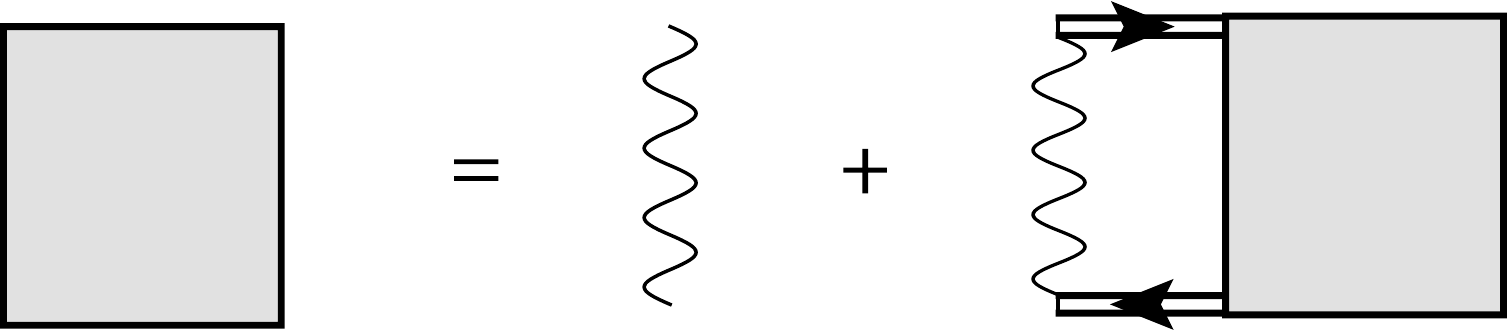}
\caption{ Recursive expression, corresponding to the Bethe-Salpeter equation for the four-point vertex.}
\label{fig:4vertex_recu}
\end{figure}

Then, the current-current correlation function obtained from Eq.~(\ref{eq:Pimnv1}) can be written as
\begin{equation}\label{eq:curcurfock}
\begin{aligned}\Big\langle J^{\mu}(x) J^{\nu}(y)\Big\rangle_{A=0}& =  \frac{\hbar c}{i} \frac{\delta\big\langle J^{\mu}(x)\big\rangle}{\delta A_{\nu}(y)}\bigg|_{A=0} \\&=
-Ne\hbar   c^2\int d^4 y' \int d^4 y''\text{tr}\bigg[\gamma^{\mu} \, G_{\psi}[0](x,y') \,\Gamma^{\nu}(y',y'';y) \,G_{\psi}[0](y'',x)\bigg].
\end{aligned}\end{equation}
The current-current correlation function is thus given by the fermion bubble diagram with dressed propagators $G_{\psi}[0]$, one bare vertex proportional to $e\gamma^{\m}$, and one dressed vertex $\Gamma^{\nu}$, as is also shown in Fig.~\ref{fig:bubble3vertex}. We can rewrite the Dyson equation as a recursive equation for $\Gamma^{\nu}$, by taking the derivative $\delta/\delta A_{\nu}$ of Eq.~(\ref{eq:dysoncond}) and using the chain rule for functional derivatives. This yields
\begin{equation}\label{eq:Gammarec}\begin{aligned}
\Gamma^{\nu}(x,x';y) &= -\frac{e}{\hbar}\gamma^{\nu}\delta^4(x-x')\delta^4(x-y) \\
 &+ \int d^4 z \int d^4 z' \frac{\delta \Sigma_{\psi}(x,x')}{\delta G_{\psi}(z,z')} \int d^4 y' \int d^4 y''\, G_{\psi}[0](z,y')\, \Gamma^{\nu}(y',y'';y) \, G_{\psi}[0](y'',z').
\end{aligned}\end{equation}
The recursive equation for $\Gamma^{\nu}$ represents an infinite bubble sum, which is diagrammatically depicted in Fig.~\ref{fig:3vertex_recu}. The zeroth-order term in Eq.~(\ref{eq:Gammarec}) is proportional to the bare vertex, and the rest is interpreted as interaction corrections to this vertex.
It is more elegant to make the current-current correlation function explicitly symmetric in the indices $\mu$ and $\nu$, by means of a four-point vertex, as is shown diagrammatically in Fig.~\ref{fig:bubble4vertex}.
This requires knowledge of the dependence of the self-energy on the Green's function, i.e., the object $\delta \Sigma_{\psi}(x,x')/\delta G_{\psi}(z,z')$. This four-point function is known in the Fock model, it is simply given by
\begin{equation}\label{eq:selfderiv}
\frac{\delta \Sigma_{\psi}(x,x')}{\delta G_{\psi}(z,z')} =\frac{g}{i}\Delta(x-x')\delta^4(x-z)\delta^4(x'-z').
\end{equation}
The required recursive equation for this four-point function is given diagrammatically in Fig.~\ref{fig:4vertex_recu}.

We can compute the conductivity in the Fock model again in the $GG$-approximation where vertex corrections are ignored. Since the fermion propagator and self-energy are the same as the propagator in the dynamical-source model, the result is also the same, i.e., the dc conductivity scales with $T^{3-4M}$. The scaling becomes linear in the limit $M\rightarrow 1/2$. Moreover, in the Fock model vertex corrections can in principle be included because the dependence of $\Sigma_{\psi}$ on $A_{\mu}$ is exactly known as we have seen.

\subsection{Finite-coupling corrections}

In section \ref{sec:fockself} the Dyson equation from Eq.~(\ref{eq:dysonpsi}) was solved exactly in the strong-coupling limit and at zero temperature, which lead to the result in Eq.~(\ref{eq:selfv3}). Taking $A_{\mu}=0$, $\vec{k}=0$ and $\omega>0$ for simplicity, the fermion Green's function in the strong-coupling limit can thus be written as $G_{\psi}(\omega) =-1/\Sigma_{\psi}(\omega) = c/\gamma^0\tilde{g}\omega^{2M}$. Here, $\tilde{g}$ has the dimensions of frequency to the power $1-2M$, and contains the numerical prefactor of the self-energy. We would like to generalize this result to also reproduce the correct weak-coupling behavior by means of a $1/\tilde{g}$ expansion. In the absence of other scales like temperature, the exact result for $G_{\psi}(\omega)$ can only depend on the ratio $\omega/\tilde{g}\omega^{2M}$. So for general values of $\tilde{g}$, the solution to the selfconsistent Dyson equation for the Green's function can be written as
 \begin{equation}\label{eq:dyson1overg}
 G_{\psi}^{-1}(\omega) =  \frac{1}{c}\gamma^0 \tilde{g} \omega^{2M} f\left(\frac{\omega}{\tilde{g}\omega^{2M}}\right).
\end{equation}
Here, $f$ is a dimensionless function, which reduces to the correct asymptotic behavior for very small and very large $\tilde{g}$ if we write a linear Ansatz for $f$ of the form
\begin{equation*}
f(x) =1+ x f'(\infty) + d(x),
\end{equation*}
where $d(x)$ is another dimensionless function that describes by construction the deviation from the linear behavior at intermediate values of $x$. To recover the strong coupling result at $x=0$, we have $f(0)=1$ and $d(0)=0$. The free result is  obtained if  $f(x)$ asymptotes to $-x$ for $x\gg1$. So $d(\infty)=d'(\infty)=0$ and $ f'(\infty)=-1$.
With this behavior of $f$, Eq.~(\ref{eq:dyson1overg}) becomes
\begin{equation*}
G_{\psi}^{-1} = \frac{1}{c}\gamma^0\left(-\omega +\tilde{g}\omega^{2M}  + \tilde{g}\omega^{2M} d\left(\frac{\omega}{\tilde{g}\omega^{2M}}\right)\right).
\end{equation*}
In the approximation where the intermediate function $d$ is ignored, we obtain
\begin{equation*}
G_{\psi}^{-1}(\omega) = -\frac{1}{c}\gamma^0\left(\omega  - \tilde{g} \omega^{2M}\right).
\end{equation*}
This Green's function is exactly a solution to the Dyson equation for very large and very small $\tilde{g}$. The kinetic term linear in $\omega$ comes about from a  $1/\tilde{g}$ expansion of the strong-coupling result. The approximation here is that this $1/\tilde{g}$ term, which is exact for vanishing $\tilde{g}$, is generalized to intermediate values of the coupling, by ignoring the deviation function $d$.
It is therefore an asystematic approximation to a systematic $1/\tilde{g}$ expansion, that nevertheless recovers the exact ultraviolet and infrared behavior of the propagator.

This form of the propagator coincides with the Green's function of the dynamical-source model from Eq.~(\ref{eq:Gretds}). The $\tilde{g}$ denotes the prefactor of the self-energy in Eq.~(\ref{eq:selfv0}).  This leads to a second interpretation of the dynamical-source model. The kinetic term that makes the $\chi$ fermion dynamical, can be seen as a phenomenological correction from the infinite coupling limit.

\section{Conclusion}\label{sec:discussion}
The main motivation of this work is to obtain a more intuitive understanding of the results of bottom-up holography and in particular the dynamical-source model. In this work we have investigated two ways to interpret the holographic results, by means of the probe-fermion model and of the Fock model. Both reproduce the fermionic self-energy in the large-$N$ limit. Firstly, the probe-fermion model gives a semiholographic interpretation of the dynamical-source model, where a dynamical probe fermion is coupled to a ``conformal field theory'', i.e., for our purposes, a theory that is scale invariant in the large-$N$ limit. The conductivity then consists of three contributions, i.e., contributions of the probe fermion, the conformal field theory and an interference term. The conformal field theory contribution consists of an $\mathcal{O}(N)$ and an $\mathcal{O}(1)$ part, the other two contributions are $\mathcal{O}(1)$. These contributions scale all three with a different power in temperature in the dc limit, which in particular makes the probe-fermion contribution experimentally distinguishable, even though it is subleading with respect to the conformal field theory.
Unfavorable properties of the probe-fermion model are that the conformal field theory becomes free in the large-$N$ limit, and that removing the anomalous dimension does not reproduce the $M=1/2$ result as anticipated. In contrast, in the Fock model, the fermionic field is itself part of the conformal field theory, and the noninteracting part of its propagator can be understood to be a phenomenological way to incorporate finite-coupling corrections. This latter interpretation also suggests that the dynamical-source model can be effectively understood in a bottom-up holographic sense.
The conductivity corresponds to the result in the dynamical-source model if vertex corrections are ignored. The latter can in principle be taken into account in the Fock model.

This work can be extended in various ways. A simple generalization would be to turn on temperature and chemical potential in the field theory, and to look at nonrelativistic cases, e.g. by considering Galilean invariance or a dynamical exponent $z$ unequal to 1. These would be the first steps towards describing the system of  cold atoms at unitarity mentioned in the introduction. There are a myriad occurrences of bottom-up models with nonzero temperature and chemical potential to compare with in the literature, see e.g. \cite{kiritsis10,SSLee09,Zaanen09,tavanfar10,zingg11,Faulkner13} for an incomplete list.
The nonrelativistic case $z\neq1$ can be studied by using a Lifshitz geometry \cite{thorlacius09,tarrio11}. It is possible to consider these cases also in semiholography, and compare them to the field-theory model. The thought behind these generalizations is to see if a simple field-theory analogy continues to hold in a more general case. The choices that we have to make in order to retrieve the holographic or semiholographic results, provide a means to interpret the latter in a more conventional, field-theoretic sense.
Then, we can have some confidence that such a field-theory construction also works the other way around, by suggesting the appropriate way to construct a (semi)holographic bottom-up model in cases where it is not immediately known or obvious. Eventually, we hope to achieve a phenomenological bottom-up dictionary, in particular one which also works for semiholography.

\acknowledgments
We thank Umut G\"ursoy, Panos Betzios and Stefan Vandoren for stimulating and relevant discussions and for critically reading the manuscript.

\appendix

\section{Notation and conventions}\label{app:conventions}

We work in SI units, unless stated otherwise. The metric tensor is given by
\begin{equation*}
\eta_{\mu\nu}~=~\text{diag}(-1,1,1,1),
\end{equation*}
 where Greek indices run over $0,1,2,3$. Latin indices run over spatial directions $1,2,3$. The spacetime position vector and four-derivative operator are respectively given by
$x^{\mu}=(ct,\vec{x})$ and $\dau_{\mu}=(1/c\;\dau_t,\vec{\nabla})$. The wave-four-vector is given by $k^{\mu}=(\omega/c,\vec{k})$. Fourier transformations to momentum space, and inverse Fourier transformations, are respectively performed as follows,
\begin{equation*}
f(x) = \int \frac{d^4 k}{(2\pi)^4} f(k) e^{ik\cdot x},\qquad f(k) = \int d^4 x f(x) e^{-ik\cdot x},
\end{equation*}
where $k\cdot x=k_{\m}x^{\mu}$ .
The Dirac matrices obey the Clifford algebra
\begin{equation*}
\left\{\gamma^{\mu},\gamma^{\nu}\right\} = 2\eta^{\mu\nu}\unit,
 \end{equation*}
 where $\unit$ denotes the 4$\times$4 identity matrix, and we note explicitly that $(\gamma^0)^2 = -\unit$.

The notation for fermionic Green's functions is as follows. The noninteracting Green's functions of the $\psi_i$ and $\chi$ fermions at $A_{\mu}=0$, appearing in sections \ref{sec:pfmodel} and \ref{sec:fock} are time-ordered connected Green's functions associated with the differential operator $ -i\slashed{\dau}$. This is done in order to keep the notation concise and Lorentz invariance manifest during the manipulations in section \ref{sec:pfmodel} and \ref{sec:fock}.
In contrast to this, in Ref.~\cite{Dirac14}, the noninteracting Green's function of the dynamical-source fermion at $A_{\mu}=0$ is the retarded Green's function corresponding to the differential operator $-i\gamma^0\slashed{\dau}$, so it includes the factor $\gamma^0$, peeled off from $\bar{\Psi}$. The reason for this definition in Ref.~\cite{Dirac14} was that this gives immediately the correct retarded Green's function that satisfies the sum rule from Eq.~(\ref{eq:sumrule}).

\section{Supersymmetric Yang-Mills theory}\label{app:sym}

For the readers without a background in supersymmetry, we briefly review the basic properties of (supersymmetric) Yang-Mills theory \cite{Proeyen12}.
Ordinary Yang-Mills theory is a non-Abelian gauge theory with gauge group $SU(N_c)$, which contains vector bosons $A^a_{\mu}$ called gluons and transforming in the adjoint representation of this gauge group.
In supersymmetric Yang-Mills theory, the symmetry is extended by adding $\mathcal{N}$ extra fermionic supersymmetry generators. These act as ladder operators on the gluon states, each time changing the spin by $\pm 1/2$, thus creating a supermultiplet of particle states. For example, in the case of global $\mathcal{N}=4$ supersymmetry, the supermultiplet consists of a single vector boson (gluon), two Dirac fermions (gluinos) and three complex scalars, denoted by $X$, which all transform in the adjoint representation of $SU(N_c)$. Note that the number of bosonic and fermionic degrees of freedom are both equal to 8, as required by supersymmetry. Supersymmetric Yang-Mills theory is a conformal field theory in $d=3+1$ dimensions, the symmetry being the conformal group $SO(4,2)$.

There is an additional global symmetry called R-symmetry, which says that the $\mathcal{N}$ supersymmetry generators can be interchanged without affecting the supersymmetry algebra. In 3+1 dimensions and for $\mathcal{N}=4$, the R-symmetry group is U(4). This corresponds to global U(4) transformations in the supermultiplet that conserve helicity, i.e., the gluinos or $X$ scalars are interchanged among themselves. This is important for our discussion of the electrical conductivity of supersymmetric Yang-Mills theory. When we speak about the electrical conductivity of the dual field theory, we imagine to take a global U(1) subgroup of this R-symmetry,
which defines a charge $e$ just like in Ref.~\cite{Kovtun08}. Unlike the gluinos and $X$ scalars, the gluon is a singlet under R-symmetry, so the gluon is electrically neutral according to this definition.

 Bottom-up models containing for instance the Maxwell or Dirac field in an asymptotically Anti-de Sitter background, are usually by construction very similar to the theory that is obtained from the benchmark duality between Type IIB string theory in AdS$_5 \times$S$_5$ and four-dimensional $\mathcal{N}= 4$ supersymmetric Yang-Mills theory in certain limits. Our working assumption is that the boundary conformal field theory dual to an asymptotically Anti-de Sitter background has similar field content and interaction structure as $\mathcal{N}=4$ supersymmetric Yang-Mills theory. Moreover, the boundary conformal field theory is supposed to be in the large-$N_c$ limit, where $N_c$ is the number of colors. Secondly, it is supposedly in the large `t Hooft coupling limit, where the `t Hooft coupling $\lambda_H = g_{YM}^2 N_c$ and $g_{YM}$ is the Yang-Mills coupling. It is also important to mention that supersymmetric Yang-Mills theory is a conformal field theory without an anomalous dimension of its fundamental fields, since the beta function of $g_{YM}$ is exactly zero. However, composite operators consisting of traces over multiple fundamental fields can obtain an anomalous dimension. If these operators are sourced, the theory is deformed away from supersymmetric Yang-Mills theory. In the bottom-up models considered here, the boundary field theory can obtain anomalous dimensions by adding extra bulk degrees of freedom that provide a source for these composite operators. For example, as in Ref.~\cite{ARPES12}, a bulk Dirac field with mass $M$ can be added and it sources a composite chiral operator in the boundary with an anomalous dimension depending on $M$. In addition,
  temperature introduces a scale to the boundary field theory. In holographic models, the presence of a black brane in the bulk spacetime breaks conformal invariance of the boundary by putting it in a thermal state with temperature $T$, coinciding with the Hawking temperature. At nonzero temperature, supersymmetry is broken in the boundary, but the superpartners of the gluons are still present in the theory. Moreover, temperature does not cause anomalous dimensions. As is clear from Fig.~\ref{fig:phasecrit}, raising the temperature of a conformal field theory will bring the system into the quantum critical region. This does not affect the scaling dimensions of the operators in the critical theory.

The field content and properties of supersymmetric Yang-Mills theory were for us the inspiration to write down the ingredients of the probe-fermion model corresponding to the contributions in Eq.~(\ref{eq:Stotpfmodel}).
Firstly, to us, the fields $\psi_i$, $\psib_i$ and $\varphi_i$
together resemble the field content of the fermionic sector of the conformal supersymmetric Yang-Mills theory. We refer to this part of the model as the ``conformal field theory'', despite the fact that it is not conformal, but only scale-invariant, for some values of the coupling $g_2$. It has a number of copies of fields $i=1,...,N$ that roughly mimics the large number of colors of the SU($N_c$) gauge theory.
The real scalars $\varphi_i$ are inspired by the gluons of supersymmetric Yang-Mills theory. They are obviously not the actual gluons of a non-Abelian gauge theory. Supersymmetric Yang-Mills theory has no anomalous dimension, but it is possible that self-interactions of the gluons in a more general non-Abelian gauge theory, caused by loop corrections, can effectively lead to an anomalous dimension. Motivated by this possibility, we give the scalar propagator an anomalous dimension parameterized by $\eta$.
The gluinos of supersymmetric Yang-Mills theory were our inspiration for the Dirac fermions $\psi_i$.
We consider for now only the fermionic sector, so there are no counterparts of the $X$ scalars of supersymmetric Yang-Mills theory.
The scalar field $\varphi_i$ remains neutral in the probe-fermion model. Our thought behind this is the fact that in supersymmetric Yang-Mills theory, the gluon is a singlet under R-symmetry as discussed above.

In the probe-fermion action, the coupling between the probe fermion and the conformal field theory is in the large-$N$ limit given by a term of the form $\bar{\chi} \pi$ and its conjugate, where $\pi \propto \sum_{i=1}^N\psi_i \varphi_i/N$.
This turns out to be similar to a composite operator of the form $O = \Tr\left[\Psi \Phi\right]$ in a typical boundary gauge theory, where the fields $\Psi$ and $\Phi$ are respectively an elementary fermionic and bosonic field in the adjoint representation of the gauge group $SU(N_c)$. This means they can be written as $\Phi = \sum_a t^a \Phi^{a}(x)$, where $a$ indexes the adjoint representation of $SU(N_c)$, and $t^a$ denotes its generators. Working this out using the usual normalization $\Tr[t^a t^b] =-\delta_{ab}/2$, we obtain $O \propto \sum_a\Psi^a \Phi^a$, where $a$ takes $N_c^2-1$ values. In the large-$N_c$ limit the 1 is negligible, so this suggests that $N$ in the probe-fermion model corresponds to $N_c^2$, and that $\pi$ may be related to a composite operator of the single-trace form $\Tr\left[\Psi\Phi\right]$, to which the dynamical source couples.

\section{Loop integral for the probe-fermion self-energy}\label{app:pfloop}

In this section the loop integral in Eq.~(\ref{eq:I}) for the fermionic self-energy in the probe-fermion model is computed at zero temperature. It is repeated here for clarity,
\begin{equation*}
I(k) = \int \frac{d^4 q}{(2\pi)^4} G_{\psi}(k+q) \,G_{\varphi}(q).
\end{equation*}
The inverse propagators of the $\varphi_i$ and $\psi_i$ fields are given in Eqs.~(\ref{eq:Gphix}) and (\ref{eq:Gpsix}), respectively. In momentum-space, they are given by
\begin{equation*}
G_{\varphi}(k) = \frac{1}{-k^{2-\eta}},
\end{equation*}
and
\begin{equation*}
G_{\psi}(k) = \frac{1}{\slashed{k}}.
\end{equation*}
Then the integral $I(k)$ becomes $I(k) = -\gamma^{\mu} \mathcal{I}_{\mu}(k)$, with
\begin{equation*}
\mathcal{I}_{\mu}(k) = \int \frac{d^4 q}{(2\pi)^4} \frac{k_{\mu}+q_{\mu}}{q^{2-\eta}(k+q)^2}.
\end{equation*}

We first compute the integral $\mathcal{I}_{\mu}$ for $\eta \neq 0$.
After subtraction of ultraviolet divergences for the real part, the integral is finite for $\eta>0$. The result for $\eta$ in the desired interval $-2 < \eta < 0$ is obtained by analytic continuation \cite{bollini64}.
After shifting the integration variable $q$ to $q-k$, to handle the tensor structure, we write the integrand as a derivative with respect to $k_{\mu}$. This yields
\begin{equation}\label{eq:splitloop}
\mathcal{I}_{\mu}(k) = -\frac{1}{\eta} \frac{\dau}{\dau k^{\mu}} \int \frac{d^4 q}{(2\pi)^4} \frac{1}{q^2 \big[(q-k)^2\big]^{-\frac{\eta}{2}}} + k_{\mu} \int \frac{d^4 q}{(2\pi)^4} \frac{1}{q^2 \big[(q-k)^2\big]^{1-\frac{\eta}{2}}} \equiv \mathcal{I}_{\mu}^{(1)}(k) + \mathcal{I}_{\mu}^{(2)}(k).
\end{equation}
We compute both terms separately, starting with the first term. We use the generalized Feynman trick,
\begin{equation*}
\frac{1}{A^{\alpha}B^{\beta} }= \frac{\Gamma(\alpha+\beta)}{\Gamma(\alpha)\Gamma(\beta)} \int_0^1 du \frac{u^{\alpha-1}(1-u)^{\beta-1}}{\big[u A + (1-u)B\big]^{\alpha+\beta}},
\end{equation*}
 which allows us to rewrite the angular integrations in terms of an Euler integral. Another appropriate shift in the integration variable removes the $k\cdot q$ terms, after which we obtain
\begin{equation*}\begin{aligned}
\mathcal{I}_{\mu}^{(1)}(k)  &= \frac{1}{2}\frac{\dau}{\dau k^{\mu}} \int_0^1 du \,u^{-1-\frac{\eta}{2}} \int \frac{d^4 q}{(2\pi)^4} \frac{1}{\big[q^2+u(1-u)k^2\big]^{1-\frac{\eta}{2}}} \\
&= - \left(1-\frac{\eta}{2}\right)k_{\mu} \int_0^1 du \,u^{-1-\frac{\eta}{2}} u(1-u) \int \frac{d^4 q}{(2\pi)^4} \frac{1}{\big[q^2 + u(1-u)k^2\big]^{2-\frac{\eta}{2}}}.
\end{aligned}\end{equation*}
The momentum integral results in
\begin{equation}\label{eq:dimregv1}
\int \frac{d^4q}{(2\pi)^4} \frac{1}{\big[q^2 +u(1-u)k^2\big]^{2-\frac{\eta}{2}}} = \frac{i}{(4\pi)^2}\frac{1}{-\frac{\eta}{2}\left(1-\frac{\eta}{2}\right)} \left[u\left(1-u\right)k^2\right]^{\frac{\eta}{2}}.
\end{equation}
With this result we can perform the $u$ integral, which is finite for $\eta>-4$, leading to
\begin{equation*}
\int_0^1 du \,\left(1-u\right)^{1+\frac{\eta}{2}} = \frac{1}{2+\frac{\eta}{2}}.
\end{equation*}
Finally, the first term of the integral $\mathcal{I}_{\mu}$ is obtained as
\begin{equation*}
\mathcal{I}_{\mu}^{(1)}(k)= \frac{i k_{\mu} \left(k^2\right)^{\frac{\eta}{2}}}{(4\pi)^2} \frac{1}{\frac{\eta}{2}\left(2+\frac{\eta}{2}\right)}.
\end{equation*}
For the second term $\mathcal{I}^{(2)}_{\mu}$ we apply the same steps. After Feynman's trick combined with appropriate shifts of the integration variable,
\begin{equation*}\begin{aligned}
\mathcal{I}^{(2)}_{\mu}(k)&= k_{\mu} \int \frac{d^4 q}{(2\pi)^4} \frac{1}{q^2\big[(q-k)^2\big]^{1-\frac{\eta}{2}}} \\
 &=  k_{\mu}\left(1-\frac{\eta}{2}\right)\int_0^1 du u^{-\frac{
\eta}{2}} \int \frac{d^4 q}{(2\pi)^4} \frac{1}{\big[q^2 +u(1-u)k^2\big]^{2-\frac{\eta}{2}}}.
\end{aligned}\end{equation*}
The momentum integral is again given by Eq.~(\ref{eq:dimregv1}). The $u$ integral is finite for $\eta>-2$ and results in a factor $1/(1+\eta/2)$. Finally, $\mathcal{I}^{(2)}_{\mu}$ becomes
\begin{equation*}
\mathcal{I}^{(2)}_{\mu}(k) = \frac{i k_{\mu} \left(k^2\right)^{\frac{\eta}{2}}}{(4\pi)^2} \frac{1}{-\frac{\eta}{2}\left(1+\frac{
\eta}{2}\right)}.
\end{equation*}
Then, the total integral $I$ is obtained by adding both terms, which gives
\begin{equation*}
I(k) = -\gamma^{\mu}\left(\mathcal{I}^{(1)}_{\mu}(k)+\mathcal{I}^{(2)}_{\mu}(k)\right) =\frac{i \slashed{k}\left(k^2\right)^{\frac{\eta}{2}}}{(4\pi)^2} \frac{1}{\frac{\eta}{2}\left(1+\frac{\eta}{2}\right)\left(2+\frac{\eta}{2}\right)},
\end{equation*}
for $-2<\eta<0$. This result was used to compute the self-energy of the probe fermion in section \ref{sec:selfft}.

For completeness, we also compute the integral $I$ at $\eta=0$. It has a logarithmic divergence for $\eta=0$. We show below how it can be regularized. We now have to deal with
\begin{equation*}
\mathcal{I}_{\m}(k)=\int\frac{d^4 q}{(2\pi)^4} \frac{q_{\m}}{q^2 (q-k)^2}.
\end{equation*}
To get rid of the tensor structure, a derivative is inconvenient because this leads to a logarithm in the integrand. Instead, we shift $q \rightarrow q+ k/2$ so that the denominator becomes even under $q \rightarrow - q$. Then, we can drop the odd term in the numerator. This is allowed as long as we regularize the integral, so that the contributions at infinity drop out. Using Feynman's trick once more, this leads to
\begin{equation*}
\mathcal{I}_{\mu}(k) = \frac{k_{\mu}}{2}\int_0^1 du \frac{d^4 q}{(2\pi)^4} \frac{1}{\big[q^2+u(1-u)k^2\big]^2}.
\end{equation*}
The loop integral is logarithmically divergent in 3+1 dimensions, but can be computed by analytical continuation to $d$ dimensions by  means of dimensional regularization,
\begin{equation*}
\int \frac{d^d q}{(2\pi)^d} \frac{1}{\big[q^2 +u(1-u)k^2\big]^2} = \frac{i}{(4\pi)^{\frac{d}{2}}} \Gamma\left(2-\frac{d}{2}\right) \left[u\left(1-u\right)k^2\right]^{\frac{d}{2}-2}.
\end{equation*}
Subsequently, the $u$ integral yields for $d>2$
\begin{equation*}
\int_0^1 du\, u^{\frac{d}{2}-2}\left(1-u\right)^{\frac{d}{2}-2} = \frac{2^{3-d} \sqrt{\pi} \, \Gamma\left(\frac{d}{2}-1\right)}{\Gamma\left(\frac{d-1}{2}\right)}.
\end{equation*}
Plugging this into $\mathcal{I}_{\mu}(k)$ and expanding around $d=4-\epsilon$, we obtain for the integral $I$,
\begin{equation}\label{eq:loopinteta0}\begin{aligned}
I(k)& = -\frac{i
\slashed{k}\left(k^2\right)^{\frac{d}{2}-2}}{(4\pi)^{\frac{d}{2}}} 2^{2-d}\sqrt{\pi} \frac{\Gamma\left(2-\frac{d}{2}\right) \Gamma\left(\frac{d}{2}-1\right)}{\Gamma\left(\frac{d-1}{2}\right)} \\
&\simeq -\frac{i \slashed{k}}{2(4\pi)^2} \left[\frac{2}{\epsilon} + \ln\left(\frac{1}{k^2}\right) + \ln\left(16 \pi\right) + \psi^{(0)}\left(\frac{3}{2}\right)\right],
\end{aligned}\end{equation}
where $\psi^{(0)}(z)$ is the digamma function, defined as $\psi^{(0)}(z) = \Gamma'(z)/\Gamma(z)$.
Our choice for the renormalization condition for Eq.~(\ref{eq:loopinteta0}) that introduces a scale into the logarithm is the following. We subtract from Eq.~(\ref{eq:loopinteta0}) $I(k_{(0)})$, that is, $I(k)$ evaluated at a constant and nonzero reference momentum scale $k_{(0)}$. Thus we obtain the renormalized loop integral at $\eta=0$,
\begin{equation*}
I(k) = \frac{i\slashed{k}}{2(4\pi)^2} \ln\left(\frac{k^2}{k^2_{(0)}}\right).
\end{equation*}
So $k^2_{(0)}$ is the value of $k^2$ at which the loop integral vanishes, which remains undetermined here and should ultimately be determined from experiment.

\section{Generating functional for current-current correlation function in the probe-fermion model}\label{app:lnZA}

Here we give in more detail the derivation of Eq.~(\ref{eq:dlnZ2pf}), which was diagrammatically represented in Fig.~\ref{fig:pfbubbles}.
We start from the generating functional given in Eq.~(\ref{eq:lnZ1}), repeated here for clarity,
\begin{equation}\label{eq:lnZ1copy}
\ln Z[A] = N \Tr \ln \Big(-G_{\psi}^{-1}[A]\Big) + \Tr \ln\Big(-N G_{\pi}^{-1}[A]\Big) + \Tr \ln\Big(-G^{-1}_{\chi}[A]\Big).
\end{equation}
In what follows, we first rearrange the terms inside the logarithms on the right-hand side so that $A$-dependent terms are separated from $A$-independent terms. This includes explicitly expanding all functionals of $A$ up to second order. Next, we expand the logarithms up to second order in $A$. Finally, we differentiate twice with respect to the gauge field, and write the result in the coordinate basis, which gives us the desired Eq.~(\ref{eq:dlnZ2pf}).

 It is convenient to write the propagators in representation-independent notation. The inverse $\psi_i$ propagator is given in Eq.~(\ref{eq:Gpsix}), and we write it as
\begin{equation*}\begin{aligned}
&G_{\psi}^{-1}[A] = G_{0}^{-1} -\frac{e}{\hbar}\slashed{A},\\
&G_0^{-1}(x,x')=-i\slashed{\dau}\delta^4(x-x').
\end{aligned}\end{equation*}
The inverse of the full Green's function $G^{-1}_{\chi}$ is given by the quadratic part in Eq.~(\ref{eq:Z3}), in basis-independent notation,
\begin{equation*}\begin{aligned}
&G^{-1}_{\chi}[A]=G^{-1}_{0}-\frac{e}{\hbar}\slashed{A}-\Sigma_{\chi}[A]\\
&\Sigma_{\chi}[A]= \frac{2ig_1^2}{g_2} + \left(\frac{2g_1}{\hbar g_2}\right)^2  G_{\pi}[A].
\end{aligned}\end{equation*}
We split the inverse $\pi$ propagator in an $A$-dependent and $A$-independent part,
\begin{equation*}\begin{aligned}
&G^{-1}_{\pi}[A]=\mathcal{G}^{-1}_{\pi} - \Sigma_{\pi}[A],\\
&\mathcal{G}^{-1}_{\pi} = \frac{2}{i\hbar^2 g_2},\\
&\Sigma_{\pi}[A]=-\frac{i}{\hbar^2} G^T_{\varphi} G_{\psi}[A],
\end{aligned}\end{equation*}
where $G^T_{\varphi}= G_{\varphi}(x',x)$ in the coordinate representation. At this point we can reexpress Eq.~(\ref{eq:lnZ1copy}) as
\begin{equation}\label{eq:lnZ2}
\ln Z[A] = N \Tr \ln\Big(-G_{0}^{-1} + \frac{e}{\hbar} \slashed{A} \Big) + \Tr\ln\Big(-N \mathcal{G}_{\pi}^{-1} + N \Sigma_{\pi}[A]\Big) + \Tr \ln\Big(-G_{0}^{-1} + \frac{e}{\hbar}\slashed{A}  +\Sigma_{\chi}[A]\Big).
\end{equation}
In the above expression, the capital trace $\Tr$ sums over both the spinor indices and the infinite matrix (coordinate or momentum) indices of the propagators.
We are going to expand this expression up to second order in the vector potential. As an intermediate step, we expand the self-energy terms which are functionals of $A_{\m}$ as
\begin{equation}\label{eq:Siexpand}
\Sigma_{\pi}[A] = \Sigma_{\pi}[0] + A_{\lambda}\frac{\delta \Sigma_{\pi}}{\delta A_{\lambda}}[0] + \frac{1}{2} A_{\lambda} A_{\rho} \frac{\delta^2\Sigma_{\pi}}{\delta A_{\lambda} \delta A_{\rho}}[0],
\end{equation}
and similar for $\Sigma_{\chi}[A]$.\footnote{Note that upon reinstating all coordinate dependence, expression Eq.~(\ref{eq:Siexpand}) is given by
\begin{equation*}\begin{aligned}
&\Sigma_{\pi}[A](x,x') \\
&= \Sigma_{\pi}[0](x,x') + \int d^4 y A_{\lambda}(y)\frac{\delta \Sigma_{\pi}(x,x')}{\delta A_{\lambda}(y)}[0] + \frac{1}{2} \int d^4 y \int d^4 y' A_{\lambda}(y) A_{\rho}(y') \frac{\delta^2\Sigma_{\pi}(x,x')}{\delta A_{\lambda}(y) \delta A_{\rho}(y')}[0].
\end{aligned}\end{equation*}
}
From now on, we suppress the notation $[0]$ indicating the absence of functional dependence on the gauge field. So it is implicit here that all propagators and vertices are evaluated with $A_{\mu}$ set to zero. With this, the generating functional Eq.~(\ref{eq:lnZ2}) becomes
\begin{equation*}\begin{aligned}
&\ln Z[A]  = N \Tr \ln\left[ - G_{0}^{-1} \Big(\unit-\frac{e}{\hbar}G_{0}\slashed{A}\Big)\right] \\
& + \Tr\ln\left[-N\Big(\mathcal{G}_{\pi}^{-1}-\Sigma_{\pi}\Big) \left\{\unit-\left[\mathcal{G}_{\pi}^{-1}-\Sigma_{\pi}\right]^{-1} \left(A_{\lambda}\frac{\delta \Sigma_{\pi}}{\delta A_{\lambda}} + \frac{1}{2}A_{\lambda}A_{\rho}\frac{\delta^2\Sigma_{\pi}}{\delta A_{\lambda}\delta A_{\rho}}\right)\right\}\right]\\
& + \Tr\ln\bigg[ \Big( -G_{0}^{-1}+ \Sigma_{\chi}\Big)\bigg\{ \unit-\left[G_{0}^{-1}-\Sigma_{\chi}\right]^{-1} \bigg(\frac{e}{\hbar}\slashed{A} + A_{\lambda}\frac{\delta \Sigma_{\chi}}{\delta A_{\lambda}}  + \frac{1}{2} A_{\lambda} A_{\rho} \frac{\delta^2\Sigma_{\chi}}{\delta A_{\lambda} \delta A_{\rho}}\bigg)\bigg\}\bigg].
\end{aligned}\end{equation*}
We can simplify this even further by rewriting $G_0^{-1}-\Sigma_{\chi} =G_{\chi}^{-1}$, the inverse $\chi$ propagator evaluated at $A_{\mu}=0$, and similarly for $\pi$ we write $\mathcal{G}_{\pi}^{-1}~-~\Sigma_{\pi}~=~G_{\pi}^{-1}$.
Then it is easy to expand $\ln Z$ up to second order in the gauge field. Using $\Tr\ln(AB)~=~\Tr\ln A~+~\Tr\ln B$ we can separate the gauge-field-dependent parts in the logarithms above. The factors $N$ drop out everywhere, except in the first term, to which just the free $\psi_i$ fields contribute. Performing the two derivatives explicitly, finally yields
\begin{equation}\label{eq:dlnZ2pf}\begin{aligned}
&- \frac{\delta^2  \ln Z[A]}{\delta A_{\m}(z) \delta A_{\nu}(z')} \bigg|_{A=0}= \\
&\;\quad N \left(\frac{e}{\hbar}\right)^2 \tr\Big[G_{0}(z,z')\g^{\nu} G_{0}(z',z)\g^{\m}\Big] \\
 &\;\quad +\int d^4x \int d^4x'\int d^4 x'' \int d^4 x''' \tr\Big[G_{\pi}(x,x')\frac{\delta\Sigma_{\pi}(x',x'')}{\delta A_{\m}(z)}G_{\pi}(x'',x''') \frac{\delta \Sigma_{\pi}(x''',x)}{\delta A_{\nu}(z')}\Big]\\
 &\;\quad + \int d^4 x\int d^4 x'\tr\Big[G_{\pi}(x,x') \frac{\delta^2 \Sigma_{\pi}(x',x)}{\delta A_{\m}(z)\delta A_{\nu}(z')}\Big]\\
 &\;\quad + \left(\frac{e}{\hbar}\right)^2 \tr\Big[G_{\chi}(z,z')\g^{\nu} G_{\chi}(z',z)\g^{\m}\Big]\\
 &\;\quad + \frac{e}{\hbar} \int d^4 x \int d^4x'\tr\Big[ G_{\chi}(x,z') \g^{\nu} G_{\chi}(z',x')\frac{\delta \Sigma_{\chi}(x',x)}{\delta A_{\m}(z)}\Big]\\
 &\;\quad + \frac{e}{\hbar} \int d^4 x \int d^4x'\tr\Big[ G_{\chi}(x,z) \g^{\m} G_{\chi}(z,x')\frac{\delta \Sigma_{\chi}(x',x)}{\delta A_{\nu}(z')}\Big] \\
&\;\quad  + \int d^4x \int d^4x'\int d^4 x'' \int d^4 x''' \tr\Big[G_{\chi}(x,x')\frac{\delta\Sigma_{\chi}(x',x'')}{\delta A_{\m}(z)}
G_{\chi}(x'',x''') \frac{\delta \Sigma_{\chi}(x''',x)}{\delta A_{\nu}(z')}\Big]\\
&\;\quad +  \int d^4 x\int d^4 x'\tr\Big[G_{\chi}(x,x') \frac{\delta^2 \Sigma_{\chi}(x',x)}{\delta A_{\m}(z)\delta A_{\nu}(z')}\Big].
\end{aligned}\end{equation}
The lowercase trace $\tr$ sums only over the spinor indices. All contributions to the right-hand side of Eq.~(\ref{eq:dlnZ2pf}) are of order $\mathcal{O}(1)$, except the first which is of $\mathcal{O}(N)$. The terms in Eq.~(\ref{eq:dlnZ2pf}) appear in the same order as the numbered diagrams in Fig.~\ref{fig:pfbubbles}.

\section{Loop integral for the Fock self-energy}\label{app:fockloop}

 This section provides some details for the computation of the loop integral in the expression for the Fock model self-energy, in the strong-coupling limit and at zero temperature. Plugging in the result for $G_{\psi}$ from Eq.~(\ref{eq:fockGpsi}), we write Eq.~(\ref{eq:dysonstrong}) as
\begin{equation}\label{eq:fockloop}
\frac{i}{ h^2(\eta)} \slashed{k} k^{\frac{\eta}{2}}= \gamma^{\mu}\mathcal{K}_{\mu}(k),
\end{equation}
where $\mathcal{K}_{\mu}$ is the loop integral that is the focus of this appendix, given by
\begin{equation*}
\mathcal{K}_{\mu}(k) = \int \frac{d^4 q}{(2\pi)^4} \frac{q_{\mu}}{\big[(q-k)^2\big]^{1-\frac{\eta}{2}} q^{2+\frac{\eta}{2}}}.
\end{equation*}
The power of $q^2$ in the denominator is the only difference with the momentum integral $\mathcal{I}_{\mu}$ of the probe-fermion model, computed in appendix \ref{app:pfloop}. The integral $\mathcal{K}_{\mu}$ can be computed in the same manner as $\mathcal{I}_{\mu}$, and as before we obtain the result for $\eta<0$ by analytic continuation.
As in Eq.~(\ref{eq:splitloop}), the numerator $q_{\mu}$ can be written as a $\dau/\dau k^{\mu}$ and a $k_{\mu}$ term, respectively,
\begin{equation*}
\mathcal{K}_{\mu}(k) = \mathcal{K}^{(1)}_{\mu}(k) + \mathcal{K}^{(2)}_{\mu}(k).
\end{equation*}
As an intermediate result, performing the same steps as for $\mathcal{I}^{(1)}_{\mu}$ in appendix \ref{app:pfloop} yields for the first term
\begin{equation*}
\begin{aligned}
\mathcal{K}^{(1)}_{\mu}(k)& = -\frac{1}{\eta}\frac{\dau}{\dau k^{\mu}} \int \frac{d^4 q}{(2\pi)^4} \frac{1}{\big(q^2\big)^{1+\frac{\eta}{4}} \big[(q-k)^2\big]^{-\frac{\eta}{2}}} \\
&= -\frac{i k_{\mu} k^{\frac{\eta}{2}}}{(4\pi)^2} \frac{4+\eta}{2\eta} \frac{\Gamma(-1-\frac{\eta}{4})\Gamma(1-\frac{\eta}{4}) \Gamma(2+\frac{\eta}{2})}{\Gamma(1+\frac{\eta}{4}) \Gamma(3+\frac{\eta}{4})\Gamma(-\frac{\eta}{2})}.
\end{aligned}\end{equation*}
For the second term we find
\begin{equation*}
\mathcal{K}^{(2)}_{\mu}(k) = k_{\mu} \int \frac{d^4 q}{(2\pi)^4} \frac{1}{q^{2+\frac{\eta}{2}}\big[(q-k)^2\big]^{1-\frac{\eta}{2}}}= \frac{i k_{\mu} k^\frac{\eta}{2}}{(4\pi)^2} \frac{\Gamma(1-\frac{\eta}{4}) \Gamma(1+\frac{\eta}{2}) \Gamma(-\frac{\eta}{4})}{\Gamma(1+\frac{\eta}{4}) \Gamma(1-\frac{\eta}{2})\Gamma(2+\frac{\eta}{4})}.
\end{equation*}
Adding these two results and solving Eq.~(\ref{eq:fockloop}) for $h(\eta)$ yields the required expression in Eq.~(\ref{eq:hfock}).
Again, the integral has a logarithmic divergence for $\eta=0$.
For a sensible result with $h(\eta)$ real, we must be certain that the result for $1/h^2(\eta)$ is positive. This is the case for $-2<\eta<0$.

\bibliography{bottomup}

\providecommand{\noopsort}[1]{}\providecommand{\singleletter}[1]{#1}%
\providecommand{\href}[2]{#2}\begingroup\raggedright\begin{thebibliography}{10}

\bibitem{Wolfle07}
H.~v.~L{\"o}hneysen, A.~Rosch, M.~Vojta, and P.~W{\"o}lfle, {\it Fermi-liquid
  instabilities at magnetic quantum phase transitions},  {\em Rev. Mod. Phys.}
  {\bf 79} (2007) 1015.

\bibitem{sondhi97}
S.~L. Sondhi, S.~M. Girvin, J.~P. Carini, and D.~Shahar, {\it Continuous
  quantum phase transitions},  {\em Rev. Mod. Phys.} {\bf 69} (1997) 315.

\bibitem{SachdevQPT}
S.~Sachdev. \textit{Quantum Phase Transitions} (Cambridge University Press,
  Cambridge, 1999).

\bibitem{vojta03}
M.~Vojta, {\it Quantum phase transitions},  {\em Rep. Prog. Phys.} {\bf 66}
  (2003) 2069.

\bibitem{Zwerger12}
W.~Zwerger. \textit{The BCS-BEC Crossover and the Unitary Fermi Gas}
  (Springer-Verlag Berlin Heidelberg, 2012).

\bibitem{Gubbels13}
K.~B. Gubbels and H.~T.~C. Stoof, {\it Imbalanced Fermi gases at unitarity},
  {\em Phys. Rep.} {\bf 525} (2013) 255--313.

\bibitem{Gubbels08}
K.~B. Gubbels and H.~T.~C. Stoof, {\it Renormalization group theory for the
  imbalanced Fermi gas},  {\em Phys. Rev. Lett.} {\bf 100} (2008) 140407.

\bibitem{sachdev97}
K.~Damle and S.~Sachdev, {\it Nonzero-temperature transport near quantum
  critical points},  {\em Phys. Rev. B} {\bf 56} (1997) 8714.

\bibitem{Herzog09}
C.~Herzog, {\it Lectures on holographic superfluidity and superconductivity},
  {\em J. Phys. A: Math. Theor.} {\bf 42} (2009) 343001.

\bibitem{Hartnoll09}
S.~A. Hartnoll, {\it Lectures on holographic methods for condensed matter
  physics},  {\em Class. Quantum Grav.} {\bf 26} (2009) 224002.

\bibitem{McGreevy10}
J.~McGreevy, {\it Holographic duality with a view toward many-body physics},
  {\em Adv. High Energy Phys.} {\bf 2010} (2010) 723105.

\bibitem{ARPES12}
U.~G{\"u}rsoy, E.~Plauschinn, H.~T.~C. Stoof, and S.~J.~G. Vandoren, {\it
  Holography and ARPES sum-rules},  {\em J. High Energy Phys.} {\bf 5} (2012)
  18.

\bibitem{Weyl13}
U.~G{\"u}rsoy, V.~P.~J. Jacobs, E.~Plauschinn, H.~T.~C. Stoof, and S.~J.~G.
  Vandoren, {\it Holographic models for undoped Weyl semimetals},  {\em J. High
  Energy Phys.} {\bf 4} (2013) 127.

\bibitem{Faulkner11}
T.~Faulkner and J.~Polchinski, {\it Semi-holographic Fermi liquids},  {\em J.
  High Energy Phys.} {\bf 6} (2011) 12.

\bibitem{erdmenger13}
J.~Erdmenger, C.~Hoyos, A.~O'Bannon, and J.~Wu, {\it A holographic model of the
  Kondo effect},  {\em J. High Energy Phys.} {\bf 12} (2013) 86.

\bibitem{kiritsis10}
C.~Charmousis, B.~Gout{\'e}raux, B.~S. Kim, E.~Kiritsis, and R.~Meyer, {\it
  Effective holographic theories for low-temperature condensed matter systems},
   {\em J. High Energy Phys.} {\bf 11} (2010) 151.

\bibitem{Liu98}
H.~Liu, {\it Scattering in anti-de Sitter space and operator product
  expansion},  {\em Phys. Rev. D} {\bf 60} (1999) 106005.

\bibitem{Balasubramanian99}
V.~Balasubramanian, S.~B. Giddings, and A.~Lawrence, {\it What do CFTs tell us
  about anti-de Sitter spacetimes?},  {\em J. High Energy Phys.} {\bf 3} (1999)
  1.

\bibitem{DHoker99}
E.~D'Hoker, S.~D. Mathur, A.~Matusis, and L.~Rastelli, {\it The operator
  product expansion of N=4 SYM and the 4-point functions of supergravity},
  {\em Nucl. Phys. B} {\bf 589} (2000) 38--74.

\bibitem{Heemskerk09}
I.~Heemskerk, J.~Penedones, J.~Polchinski, and J.~Sully, {\it Holography from
  conformal field theory},  {\em J. High Energy Phys.} {\bf 10} (2009) 79.

\bibitem{Fitzpatrick11}
A.~L. Fitzpatrick, E.~Katz, D.~Poland, and D.~Simmons-Duffin, {\it Effective
  conformal theory and the flat-space limit of AdS},  {\em J. High Energy
  Phys.} {\bf 7} (2011) 23.

\bibitem{Papadodimas12}
S.~El-Showk and K.~Papadodimas, {\it Emergent spacetime and holographic CFTs},
  {\em J. High Energy Phys.} {\bf 10} (2012) 106.

\bibitem{Kovtun08}
P.~Kovtun and A.~Ritz, {\it Universal conductivity and central charges},  {\em
  Phys. Rev. D} {\bf 78} (2009) 066009.

\bibitem{Dirac14}
V.~P.~J. Jacobs, S.~J.~G. Vandoren, and H.~T.~C. Stoof, {\it Holographic
  interaction effects on transport in Dirac semimetals},  {\em Phys. Rev. B}
  {\bf 90} (2014) 045108.

\bibitem{sachdev13}
Y.~Huh, P.~Strack, and S.~Sachdev, {\it Conserved current correlators of
  conformal field theories in 2 + 1 dimensions},  {\em Phys. Rev. B} {\bf 88}
  (2013) 155109.

\bibitem{umutpanos}
U.~G{\"u}rsoy and P.~Betzios. \textit{Fermionic Green's functions in an
  electromagnetic background}, work in progress.

\bibitem{SSLee09}
S.-S. Lee, {\it Non-Fermi liquid from a charged black hole: A critical Fermi
  ball},  {\em Phys. Rev. D} {\bf 79} (2009) 086006.

\bibitem{Zaanen09}
M.~{\v C}ubrovi{\'c}, J.~Zaanen, and K.~Schalm, {\it String theory, quantum
  phase transitions, and the emergent Fermi liquid},  {\em Science} {\bf 325}
  (2009) 439--444.

\bibitem{tavanfar10}
S.~A. Hartnoll and A.~Tavanfar, {\it Electron stars for holographic metallic
  criticality},  {\em Phys. Rev. D} {\bf 83} (2011) 046003.

\bibitem{zingg11}
V.~Giangreco, M.~Puletti, S.~Nowling, L.~Thorlacius, and T.~Zingg, {\it
  Holographic metals at finite temperature},  {\em J. High Energy Phys.} {\bf
  1} (2011) 117.

\bibitem{Faulkner13}
T.~Faulkner, N.~Iqbal, H.~Liu, J.~McGreevy, and D.~Vegh, {\it Charge transport
  by holographic Fermi surfaces},  {\em Phys. Rev. D} {\bf 88} (2013) 045016.

\bibitem{thorlacius09}
U.~H. Danielsson and L.~Thorlacius, {\it Black holes in asymptotically Lifshitz
  spacetime},  {\em J. High Energy Phys.} {\bf 3} (2009) 70.

\bibitem{tarrio11}
J.~Tarrio and S.~Vandoren, {\it Black holes and black branes in Lifshitz
  spacetimes},  {\em J. High Energy Phys.} {\bf 9} (2011) 17.

\bibitem{Proeyen12}
D.~Z. Freedman and A.~van Proeyen. \textit{Supergravity} (Cambridge University
  Press, Cambridge, 2012).

\bibitem{bollini64}
C.~G. Bollini, J.~J. Giambiagi, and A.~G. Dom{\'i}nguez, {\it Analytic
  regularization and the divergences of quantum field theories},  {\em Il Nuovo
  Cimento Series 10} {\bf 31} (1964) 550--561.

\end{thebibliography}\endgroup

\end{document}